\documentclass[8pt, twocolumn]{article}

\usepackage{amssymb,amsmath,amsthm,mathtools}
\usepackage[breaklinks=true]{hyperref}
\usepackage{ccicons}
\usepackage{url}
\usepackage[usenames,dvipsnames]{xcolor}
\usepackage{graphicx}
\usepackage{cleveref}
\usepackage{algorithm}
\usepackage{algpseudocode}
\usepackage{listings}
\hypersetup{
  colorlinks,
  linkcolor	=	Black,
  urlcolor 	= Black,
  citecolor	=	Black,
  pdfauthor={Mirco Richter},
  pdfkeywords={crypto, bft, bitcoin, cryptocurrency, consensus, blockchain},
  pdftitle={Crisis, Linear Order},
  pdfsubject={Crisis},
  pdfpagemode=UseNone
}

\theoremstyle{plain}
\newtheorem{theorem}{Theorem}[section]
\newtheorem{proposition}[theorem]{Proposition}

\newtheorem{corollary}[theorem]{Corollary}

\newtheorem{definition}[theorem]{Definition}

\theoremstyle{remark}
\newtheorem*{remark}{Remark}
\newtheorem{example}{Example}

\title{Crisis: Probabilistically Self Organizing Total Order in Unstructured 
P$\cdot$2$\cdot$P Networks}
\author{$\star$ Mirco Richter $\star$ \\ (mirco.richter@mailbox.org)}

\begin{document}
%******************************
% New width of multlined

%******************************
\newlength{\mylength}
\setlength{\mylength}{\linewidth-2\multlinegap}

\maketitle
\setcounter{tocdepth}{1}

%****************************************************************
%							***     Introduction ****
%****************************************************************
\newpage
\section{Introduction}
In their pioneering, but largely ignored work \textit{"{}Byzan-tine-Resistant Total Ordering Algorithms"} \cite{MM}, the authors Moser \& Melliar-Smith established total order on network events utilizing a concept, best described as \textit{virtual voting}. This simple yet ingenious insight achieves full asynchrony and close to optimal communication overhead as almost no additional information has to be send, besides the actual payload. Instead messages acknowledging other messages are just interpreted as virtual processes executing some consensus algorithm to decide the total order. 

Unfortunately, Moser \& Melliar-Smith's approach is based on a byzantine fault tolerant protocol, that inevitably depends on the number of participants, or the overall voting weight in the system. Those algorithms are therefore useless when it comes to fully local, unstructured Peer-to-Peer networks and their ever changing number of participants, or potentially unbounded voting power.

However, in \textit{"{}Byzantine Agreement, Made Trivial"} \cite{SM}, Micali described a so called player replaceable consensus protocol, that is able to execute each step of the computation inside an entirely different set of processes. \textit{Player replaceability} is therefore a real paradigm shift when it comes to agreement in open systems. In particular, it allows for previously unknown solutions to the BFT-CUP problem as described by Alchieri et al. in \cite{ABFG}.

It is true, that Micali's protocol needs some level of synchronism, but this is where the full power of virtual voting really shows up: 

Inside the virtual setting, synchronism can be simulated, while keeping the actual system fully asynchronous from the outside. In fact such a behavior can be achieved, by simply interpreting messages as clock ticks of even length, regardless of the amount of real world time it took them to arrive. We call this phenomena \textit{virtual synchrony} and point out, that it appears pretty straight forward from the internal logic of Lamport clocks. 

With all this in mind, a combination of Micali's player replaceability and Moser \& Melliar-Smith's idea of virtual voting seems quite obvious, provided the goal is total order on messages in unstructured Peer-to-Peer networks. 

The Crisis protocol family follows this line of thought and presents a framework for asynchronous, signature free, fully local and probabilistically converging total order algorithms, that may survive in high entropy, unstructured Peer-to-Peer networks with near optimal communication efficiency. 
Regarding the natural boundaries of the CAP-theorem, Crisis chooses different compromises for consistency and availability, depending on the severity of the attack.

The family is parameterized by a few constants and external functions called voting-weight, incentivation \& punishement, difficulty oracle and quorum-selector. These functions are necessary to fine tune the dynamics and very different long term behavior might appear, depending on any actual choice. Since proper function design is highly important, Cisis should be seen more as a total order framework, than as an actual algorithm. 

\subsection*{Historical note}
It is a concerning and somewhat wired incident in history that no reference to the foundational  work of Moser \& Melliar-Smith \cite{MM} seems to appear in any of the modern literature on virtual voting based approaches like hashgraph \cite{BL}, parsec \cite{CKHMS}, or blockmania \cite{DH} at the time of this writing. In contrast, some of the later references make it look like virtual voting is a new invention. However it is neither new, nor is it an invention at all: 

To the best of the authors knowledge, the ability to execute agreement protocols 'virtually' (by which we mean that "{}[..] \textit{votes are not contained explicitly in the messages, but are deduced from the causal relationships between messages} [..]" \cite[p.84]{MM}), was first observed by Moser \& Melliar-Smith in 1993. The authors showed, that such a property is inherent in any so called byzantine partial order, which are more or less just cryptographically secured Lamport timestamps, that appear naturally whenever a message acknowledges another message. Virtual voting is therefore not a new invention, but an \textit{emergent mathematical phenomena} on certain graphs, that have a structure similar to the one used in the Lamport clocks.
\begin{remark}
This strange glitch in time might exist, at least partially, due to monetary interest. It shines a scary light on the fragility of scientific standards, when huge econimic interest are suddenly involved. The antidote is proper education and any reader interested in historical correctness is encouraged to look at the origins of those ideas as started by Lamport \cite{LL} back in 1978. 

After all, the work of Moser \& Melliar-Smith \cite{MM} deserves all the credits, when it comes to virtual voting on byzantine partial order graphs. That paper has the potential to be seminal, despite the fact that it appears not referenced at all in most modern approaches, at the time of this writing.
\end{remark}
\subsection*{License} This work is published under version 4.0 of the creative commons license \ccbyncsa. For commercial use, please contact the author.

\subsection*{Donations}
If you would like to support the continuous production of content like this, please donate via one of the following channels, or contact the author for additional solutions:\\
Bitcoin:\\ 
1B5DNwRGC3Kb2MbPuJB4cQX9UsChPUvUWf\\
PayPal: mirco.richter@mailbox.org

\section{Model of Computation} 
\subsection{Random oracle model} We work in the random oracle model, e.g. we assume the existence of a cryptographic hash function, that behaves like a random oracle. We write 
\begin{equation}
H: \{0,1\}^*\to \{0,1\}^p
\end{equation}
for such a function, as it maps binary strings of arbitrary length onto binary strings of fixed length $p$. As usual, we call $H(b)$ the \textit{digest} value of the binary string $b$ and assume $H$ to be collision-, preimage- and second preimage-resistant.

\subsection{Network model} The protocol is executed in a dynamic, distributed system, where processes might join or leave at any time. We therefore have to assume some sort of simple message-oriented transport protocol, such that each participating processes is eventually able to send or receive data packages. In addition, the system is considered fully asynchronous in that no bound can be placed on the time required for a computation or for communication of any message.

A process is called \textit{honest} at time $t$, if it executes the protocol according to the rules at that time and it is called \textit{faulty} if it deviates from the protocol in one way or another. 

\section{Data Structures}
\subsection{Messages} Messages distribute payload across the network and the purpose of the present paper is to establishe a total order on those messages, that respects causality and is probabilistically invariant among all honest participants. 

The system is open and anyone is able to inject an arbitrary amount of messages at any given moment in time. However we crucially require the existence of a function, that assigns a weight factor to any such message. The purpose of this weight is both to prevent Sybil and system scale DOS attacks and to provide any message with a certain amount of voting power to influence the generated order. 

Proper weight function design is therefore of major importance when it comes to behavior control and self-organization. Different dynamics might appear relative to any weight function, some of which are stable and some of which are not.

To define our message type, we expand the ideas of Lamport \cite{LL} as well as Moser \& Melliar-Smith \cite{MM} and use additional insight from blockmania \cite{DH} and Bitcoin \cite{SN}. Giving three fixed protocol constants $c1,c2,c3\in \mathbb{N}$, a message is then nothing but a byte string of variable length, subject to the following interpretation:

{\scriptsize
\begin{lstlisting}[label=struct:message]
struct Message{
    byte[c1] nonce,
    byte[c2] id,
    byte[c3] num_digests,
    byte[p * num_digests] digests,
    byte[ ] payload
}
\end{lstlisting}
}
In this definition, the $nonce$ is a general purpose byte field of fixed length. It might be required to compute the weight function in an actual incarnation of the protocol. For example, if a protocol weight function is similar to Hashcashs Proof-of-Work \cite{SN}, the nonce is necessary to probe the search space of hash values. If, on the other hand, the protocol uses a Proof-of-Stake or Proof-of-Authority style weight function, the nonce might contain a signature of the message to verify ownership of some staked voting weight.

In addition, payload can be anything, that is properly serializeable into a bytefield. Other then that, Crisis makes no assumptions on its internal structure. In any case, the outcome of the protocol is a total order, e.g. a chain of payload chunks.

Moreover, $id$ is a binary string used to group messages into what we call virtual processes. Its neither a unique identifier of a message, nor must it represent an actual process. Its main purpose is to clearify how the ideas of BFT-CUP \cite{ABFG} emerge in our virtualized setting. The last messages of a virtual round with an identical $id$ will be considered as votes from the virtual process. However, depending on the weight function, it might be possible that different real world processes collaborate under the same virtual process $id$.

The $num\_digests$ field is just a standard way to represent the length of the following byte array $digests$, the latter of which contains digest values that acknowledge the existence of other messages, or the empty string, in case the message does not acknowledge any other message\footnote{Acknowledgement of the empty string is straight forward and easily definable as the hash of the empty string $H(\{\})$.}. We assume, that $digests$ contains any digest only once, which implies that we work with graphs not with multi-graphs, later on. 

The key insight here is, that a message that acknowlede other messages defines an inherent natural causality. To the best of the authors knowledge, this by now standard mechanism was derived in great detail by Lamport in his paper \cite{LL} from 1978 and we encourage the interested reader to look at the original source for further explanations. 

In any case, if $m$ and $\acute{m}$ are two messages, we write 
\begin{equation}
\label{def:knowledge-arrow}
m\to \acute{m}\,,
\end{equation}
if and only $m$ acknowledges $\acute{m}$, that is the digest $H(\acute{m})$ of $\acute{m}$ is contained in the field $m.digests$. We then say that $m$ is a \textbf{direct effect} of $\acute{m}$, or that $\acute{m}$ is a \textbf{direct cause} of $m$ and that both are in a direct causal relation\footnote{We chose this arrow convention to be more in line with the ideas of BFT-CUP \cite{ABFG}. The arrow can be interpreted as "has knowledge of".}.

In what follows, we write \textsc{message} for the set of all messages and postulate a special non-message $\oslash\in\textsc{message}$\footnote{In what follows, this message will indicate the inability of the system to agree on any actual message in a given voting period.}. Moreover we assume the existence of a string metric $d: \textsc{message} \times \textsc{message} \to \mathbb{R}$ like the Levenshtein distance, such that $(\textsc{message},d)$ is a metric space and we are able to talk about the distance $d(m,\acute{m})$ between two messages.

\subsubsection{Weight systems}
\label{sec:weight-system} The protocol assumes the existence of a so called weight system, which assigns a certain value to any given message and defines a way to combine the weight of different messages. It also provides a minimum threshold on message weight for the prevention of Sybil attacks. The choice of such a system is crucial and the overall dynamic of the system depend on it.
\begin{definition}[Weight system] Let $(\textsc{message},d)$ be the metric space of all messages and $(\mathbb{W},\leq)$ a totally ordered set. Then the tuple $(\mathbb{W}, w,\oplus, c_{min})$ is a \textbf{weight system}, if $w$ is a function
\begin{equation}
\label{def:weight-function}
w: \textsc{message} \to \mathbb{W}
\end{equation}
that assigns a an element of $\mathbb{W}$ to any message, called the \textbf{weight function}, $c_{min}\in \mathbb{W}$ is a constant, called the \textbf{weight threshold} and $\oplus$ is a function
\begin{equation}
\label{def:weight-sum}
\oplus: \mathbb{W} \times \mathbb{W} \to \mathbb{W}
\end{equation}
called the \textbf{weight sum}, such that the following characteristic properties are satisfied:\\
-- Tamper proof: Let $m\in\textsc{message}$ be a message, with weight $w(m)\geq c_{min}$ and let $\acute{m}\neq m$ be another message, close to $m$ in the metric $d$. Then $w(\acute{m})< c_{min}$, with high probability.\\
-- Uniqueness: If there are two messages $m$ and $\acute{m}$ with $m\neq\acute{m}$, then  $w(m)\neq w(\acute{m})$ with high probability. \\
-- Summability: $(\mathbb{W},\oplus)$ is a totally ordered, abelian group.
\end{definition}
\begin{remark}
If $(\mathbb{W}, w,\oplus, c_{min})$ is a weight system, we sometimes write 
$\ominus\, x$ to indicate the inverse of an element $x$ in the group $(\mathbb{W},\oplus)$ and $x\ominus y$ for the sum with such an inverse. Moreover, if $M$ a set of messages, we write
\begin{equation}
\label{def:overall-weight}
w(M):= \textstyle\bigoplus_{m\in M} w(m)
\end{equation}
for the sum of the individual weights of all messages from $M$ and call it the (overall) weight of $M$. In addition we use the convention $w(\emptyset) = 0$, where $0$ is the neutral element in $\mathbb{W}$. 
\end{remark}

Given any message $m\in \textsc{message}$, the value $w(m)$ is interpreted as the amount of voting power, $m$ holds to influence total order generation. The temper proof property  assures, that processes can not change messages easily, without dropping their weight below a certain threshold. As explained by Beck in his 2002 paper \cite{BA} on Hashcash, such an approach ensures resistance against Sybil and certain DOS attacks without the need for any Signature scheme. It is famously utilized in the Nakamoto consensus family \cite{SN}.

However in contrast to Nakamoto consensus, the present protocols are leaderless and voting is a collective process, where the overall voting weight is a combination of individual weights. The system therefore needs a way to actually execute this combination. This is reflected in the weight sum operation $\oplus$\footnote{Weight systems might use ordinary addition or multiplication as their weight sum definition, however other ways to combine individual weights might be more realistic in certain setups.}. 

\subsubsection{Causality}
\label{sec:causality} Messages may contain digests of other messages, which in turn contain digest of yet other messages and so on. This represents quite literally a partial order of causality: For a message $m$, to incorporate an acknowledgement of another message $\acute{m}$, message $\acute{m}$ must have existed before $m$, which implies that we can talk about the past and the future of any given massage. However a message might neither be in the past nor in the future of another message and those 'spacelike' messages are therefore not comparable. The purpose of a total order algorithm is then to extend the causal order into a total order, such that all messages become comparable. 

To the best of the authors knowledge, this natural idea appeared for the first time in 1978 as part of Lamports seminal paper \cite{LL} under the term happens-before relation. Another frequently used term is 'spacetime' diagram, because the causal partial order between messages behaves very much like a spacetime diagram in special relativity. It is famously used in Lamport timestamps and was later adopted by Moser \& Melliar-Smith, as foundation for what we might now call \textit{virtual agreement} or virtual voting\footnote{Much later, algorithms like hashgraph \cite{BL}, parsec \cite{CKHMS}, or blockmania \cite{DH} adopted this in one way or another, unfortunately without any reference to the original ideas.}. The following definition provides our incarnation of Lamports original ideas, adopted to our messages type:

\begin{definition}[Causality]
\label{def:causal-relation}
Let $m,\acute{m}\in \textsc{message}$ be two messages. Then $\acute{m}$ is said to \textbf{happen before} $m$, if $m=\acute{m}$ or if there is a (possibly empty) sequence of messages $m_1, \cdots, m_k$, such that  $m\to m_k \to \cdots \to m_1 \to \acute{m}$. In that case we write $\acute{m}\leq m$, call $m$ an \textbf{effect} of $\acute{m}$ and $\acute{m}$ an \textbf{cause} of $m$ and say that there is a \textbf{causality chain} from $\acute{m}$ to $m$.

Comparable messages are moreover called \textbf{timelike}, while incomparable messages are called \textbf{spacelike}. If messages $\acute{m}$ and $m$ are timelike, $\acute{m}$ is said to be in the \textbf{past} of $m$ and $m$ is said to be in the \textbf{future} of $\acute{m}$, if $\acute{m}\leq m$.
\end{definition}

\subsubsection{Vertices}
\label{sec:Vertex}  
To establish our total order, messages have to be extended by a small amount of local voting data, that is not transmitted to other processes. In fact, no votes are send through the network at all, but are deduced from the causal relation between messages. This is a key characteristic of virtual voting based systems, explicitly stated by Moser \& Melliar-Smith in \cite{MM}. We call such an extension a \textit{vertex}:

{\scriptsize
\begin{lstlisting}[escapeinside={(*}{*)}]
struct Vertex{
    Message m,
    Option<uint> round,
    Option<boolean> is_last,
    Option<TotalOrderSet<uint>> svp ,
    Option<(Message,Option<boolean>)>[ ] vote,
    Option<uint> total_position
}
\end{lstlisting}
}
We write \textsc{vertex} for the set of all vertices and assume that any entry of option type is initialized with the default value, which we symbolize as $\bot$. Properties of messages are then easily extended to corresponding properties of vertices and we write:
\begin{equation}
\scriptsize
\label{eq:equal-vertex-properties}
\begin{array}{l}
w(v)\leftarrow w(v.m),\\
v.nonce \leftarrow v.m.nonce,\\
v.id \leftarrow v.m.id,\\
v.num\_digests \leftarrow v.m.num\_digests,\\
v.digests \leftarrow v.m.digests,\\
v.payload \leftarrow v.m.payload
\end{array}
\end{equation}
If $v$ is a vertex, $v.m$ is called the \textit{underlying} message of $v$. It is important to note, that equal messages might not result in equal vertices, as the appropriate vertices might have otherwise different entries. We therefore have to loosen the rigidity of equality a bit and use the following definition of equivalence instead.
\begin{definition}[Equivalence of vertices]
\label{def:vertex-equality}
Let $v$ and $\acute{v}$ be two vertices with equal underlying messages, i.e. $v.m=\acute{v}.m$. Then $v$ and $\acute{v}$ are said to be \textbf{equivalent} and we write 
$v\equiv\acute{v}$.
\end{definition}
The causal relation (\ref{def:causal-relation}) between messages can then be extended to a causal relation between vertices. 
\begin{definition}[Vertex causality]
\label{def:vertex-causal-relation}
Let $v,\acute{v}\in \textsc{Vertex}$ be two vertices. Then $\acute{v}$ is said to happen before $v$, iff $\acute{v}.m\leq v.m$. In that case we call $v$ an effect of $\acute{v}$ and $\acute{v}$ an cause of $v$ and say that there is a causality chain from $\acute{v}$ to $v$. Comparable vertices are moreover called \textbf{timelike}, while incomparable vertices are called \textbf{spacelike}. If vertices $\acute{v}$ and $v$ are timelike, $\acute{v}$ is said to be in the \textbf{past} of $v$ and $v$ is said to be in the \textbf{future} of $\acute{v}$, if and only if $\acute{v}\leq v$.
\end{definition}

\subsection{Lamport graphs} 
As partially ordered sets are more or less the same thing as directed acyclic graphs by the categorical $\textsc{dag} \models \textsc{poset}$ adjunction \cite[sec 5.1]{SD}, sets of causaly ordered vertices have a natural graph structure, which we call   a Lamport graph. As implicitly understood by Moser \& Melliar-Smith \cite{MM}, those graphs are well suited for the generation of total order on network events.

Nevertheless, care must be taken when it comes to an actual set of vertices, as such a set might not be ordered at all, if it contains a vertex without all its acknowledging vertices. This motivates our definition of Lamport graphs as a vertex set, closed under the causality relation:  
\begin{definition}[Lamport Graph]
Let $V\subset \textsc{Vertex}$ be a finite set of vertices, such that $V$ contains all vertices $\acute{v}$ with $\acute{v}\leq v$ for all $v\in V$, but no two vertices in $V$ are equivalent. Then the graph $G=(V,A)$ with $(v,\acute{v})\in A$, if and only if $v\to\acute{v}$ is called a \textbf{Lamport graph}. Moreover, if $v$ is a vertex in a Lamport graph $G$, the subgraph $G_v$ of $G$ that contains all causes of $v$ is called the \textbf{past} of $v$.

Two Lamport graphs are said to be \textbf{equivalent}, if they are isomorphic as graphs and their vertex sets are equivalent, that is every vertex in one graph has an equivalent vertex in the other  and vice versa.
\end{definition}
Lamport graphs are directed and  acyclic for all practical purposes, because the inducing causality relation (\ref{def:vertex-causal-relation}) between vertices is a partial order, with very high probability. The proof of the following proposition makes this precise.
\begin{proposition}Let $G$ be a Lamport graph. Then, for all practical purposes, $G$ is directed and acyclic.
\end{proposition} 
\begin{proof}The proof is based on the assumption, that our hash function practically prevents causality loops, in other words, it is infeasible to generate vertices $v_1, \ldots,v_k$, such that $v_1 \to v_2 \to \cdots \to v_k$, but $v_1 = v_k$ for some $k\geq 2$. Under this assumption,  definition (\ref{def:vertex-causal-relation}) provides a partial order on a vertex set $V$ and the proposition follows from the categorical adjunction between posets and directed acyclic graphs.

To see the partial order on $V$ in detail, first observe that reflexivity is immediate, since any vertex causally follows itself by definition (\ref{def:vertex-causal-relation}). Transitivity is deduced from (\ref{def:vertex-causal-relation})  in a similar fashion, as $v\leq \acute{v}$ and $\acute{v} \leq \tilde{v}$ implies the existence of causal chains $\acute{v}\to v_k \to \cdots \to v_1 \to v$ and $\tilde{v}\to w_j \to \cdots \to w_1 \to \acute{v}$, which combine into a causal chain from $\tilde{v}$ to $v$, hence $v\leq\tilde{v}$.

We proof antisymmetry by contradiction and assume $v\neq \acute{v}$, but $v\leq \acute{v}$ as well as $\acute{v}\leq v$. Then there are causal chains $\acute{v}\to v_k \to \cdots \to v_1 \to v$ and $v\to w_j \to \cdots \to w_1 \to \acute{v}$, which implies that there is a causal chain loop $v\to w_j \to \cdots \to v_1 \to v$. This however violates our assumption on the infeasibility of generating those loops. 
\end{proof}
Since the geometric structure of a Lamport graph is fully determined by its underlying set of messages, the past of equivalent vertices is the same in any graph. This key feature is crucial in the generation of an invariant total order and the following key theorem makes this precise.
\begin{theorem}[Invariance of the past] 
\label{thm:invariant-past}
Let $v\in G$ and $\acute{v}\in\acute{G}$ be two equivalent vertices in two Lamport graphs. Then the past of $v$ in $G$ is a Lamport graph, equivalent to the past of $\acute{v}$ in $\acute{G}$, for all practical purposes and $G_v$ and $\acute{G}_{\acute{v}}$ have equal cardinality, i.e. $|G_v|=|\acute{G}_{\acute{v}}|$.
\end{theorem}
\begin{proof}Recall that the cardinality of a finite graph is equal to the number of its vertices. We start our proof with the simple observation, that the past of a vertex in a Lamport graph is a Lamport graph, since it trivially contains all elements from its past. It therefore remains to show, that the vertex sets from $G_v$ and $\acute{G}_{\acute{v}}$ are equivalent and of equal size.

To see that, first observe that $v.digests$ and $\acute{v}.digests$ actually contain the same digests, as $v.m=\acute{v}.m$ follows from our definition of equivalence (\ref{def:vertex-equality}). Since $G$ is a Lamport graph, it must contain a set of $v$'s direct causes $S_v:=\{x\in G\;|\; H(x.m)\in v.digests\}$ and since $\acute{G}$ is a Lamport graph too, it must also contain a set of $\acute{v}$'s direct causes $S_{\acute{v}}:=\{y\in \acute{G}\;|\; H(y.m)\in \acute{v}.digests\}$. However, since $H$ is a cryptographic hash function, we know that $x.m$ is equal to $y.m$ with very high probability for all $x\in S_v$ and $y\in S_{\acute{v}}$, due to the second preimage resistance of our hash function $H$. This implies that all vertices in $S_v$ and $S_{\acute{v}}$ are equivalent, with very high probability. Moreover $S_v$ and $S_{\acute{v}}$ are of equal size, since no Lamport graph contains equivalent vertices.

The same argument can then be applied to all pairwise equivalent vertices $x\in S_v$ and $\acute{x}\in \acute{S}_{\acute{v}}$ with $x \equiv \acute{x}$, which proofs the proposition by induction, since both $G_v$ and $\acute{G}_{\acute{v}}$ are finite.
\end{proof}
\begin{remark}
In our incarnation, Lamport graphs do not necessarily represent actual network communication. All they represent is causal order between messages. Crisis therefore allows for 'ghost processes', which just route \& distribute data without ever generating messages themselves. Those processes are entirely transparent from the inside of any Lamport graph and are forbidden per definition in 'gossip-over-gossip' style adaptations of Lamports original ideas as used in \cite{BL}, or \cite{CKHMS}. We believe that our approach is more general and works better under sophisticated byzantine behavior in fully local and unstructured Peer-2-Peer networks.
\end{remark}

\section{Communication} Crisis is build on top of two simple push\&pull gossip protocols, that are used for the distribution of messages and to keep local knowledge of neighbors up to date. Such gossip algorithms are well suited for the communication in unstructured Peer-2-Peer networks, as seen in real world applications like Bitcoin. However, a developer is free to choose any other approach, if necessary. All the system needs, is a way to distribute messages in a byzantine prone environment.

\subsection{Message generation} All network communication starts with the generation of messages which are then distributed using the protocols delivery system. However messages must satisfy a certain structure to be redistributed by any honest process. This is an effective first measure against the easily detectable part of faulty behavior. Algorithm (\ref{alg:message-integrity}) shows how a honest process generates a valid message $m$, assuming that $nonce$ is chosen in such a way, that $w(m)>c_{min}$:

\begin{algorithm}
\scriptsize
\label{alg:message-generation}
\begin{algorithmic}[1]
\Procedure{message}{id, nonce:n, load:p, lamport\_graph:G}
\State Find a last vertex $v$ with $v.id=id$ in $G$
\State Choose $\acute{S}\subset \{\acute{v}.m\;|\; \acute{v}\in G\,\wedge\, \acute{v}\not\in G_v \}$, such that 
\Statex \quad\, all elements of $\acute{S}$ have different $id$'s
\State \Return [n, id, $|\acute{S}\cup\{v.m\}|$, 
$\{H(\acute{m})\;|\; \acute{m}\in \acute{S}\cup\{v.m\}\}$, p]
\EndProcedure 
\end{algorithmic}
\caption{Generate message}
\end{algorithm}

According to algorithm (\ref{alg:message-integrity}), a honest process generates a message by including a digest of the last message that it knows with the same $id$. A message is called a last message of a given id, if it is not in the past of any other message under the same id. In addition a set of digests from messages is incorporated, that are not in the past of the already included last message with the same $id$. This latter set is otherwise undetermined by the protocol and any choice is valid. 
\begin{remark}
In an actual application, a honest process might just incorporate acknowledgements to a random subset of all messages that it received after the generation of the previous message $m$ with the same $id$. Those messages can not be in the past of $m$, due to the definition of Lamport graphs and are therefore valid. However the same process might as well apply a more sophisticated strategy for the inclusion of messages, depending on the incentivation and punishment strategy of the system. 
\end{remark}

If a valid message is generated, the appropriate process should generate an new vertex and write it into its own Lamport graph for further distribution. The following section describes the proper way to handle this situation.

\subsection{Lamport graph extension} After some process obtains a byte string that might be a message, it has to rule out all immediately observable faulty behavior and then check the integrity of that message against its own Lamport graph. If everything works out, the Lamport graph is extended with a vertex including the new message, if not, the message is deleted.

If the message is not already known, the procedure starts with a low level check against the basic structure of a massage, including bound checks and things like that. We abstract this as a boolean valued function \textsc{bytelevel\_correctness}. After that, the process checks the weight of the message to see if it is above the minimum threshold bound $c_{min}$. To check the payload, we assume the existence of a boolean function \textsc{payload\_correctness} that compares the payload against the system rules. 

If all this works out properly, the process checks the entries in $m$'s field of digests $m.digests$. All referenced messages must have exactly one corresponding vertex in the current Lamport graph and all of theses vertices must have different $id$'s. The process then looks for vertices with the same id as the one in the message. If there are some in the current Lamport graph, the process makes sure that one of these messages is referenced in $m.digests$.

If any of this does not work out, the message is considered faulty and is deleted. If on the other hand, everything is ok, the Lamport graph is extended with a new vertex that contains the message and new edges that points from the vertex to all vertices with messages referenced in $m.digests$. Algorithm (\ref{alg:message-integrity}) shows the details:
\begin{algorithm}
\caption{Message integrity}
\label{alg:message-integrity}
\scriptsize
\begin{algorithmic}[1]
\Procedure{integrity}{message:m, lamport\_graph:G}
\State \textbf{if} $\textsc{bytelevel\_correctness}(m)$ \textbf{and}
\Statex \quad\,\quad\, $w(m)> c_{min}$ \textbf{and}
\Statex \quad\,\quad\, $\textsc{payload\_correctness}(m.payload)$ \textbf{and}
\Statex \quad\,\quad\, there is no vertex $v\in G$, with $v.m=m$ \textbf{and}\label{alg-2:line10}
\Statex \quad\,\quad\, every $H\in m.digests$ references a vertex in $G$ \textbf{and}
\Statex \quad\,\quad\, all referenced vertices have different $id$'s
\State \textbf{then}
\State \quad\,\textbf{if} there is a vertex $v\in G$ with $v.id=m.id$ \textbf{then}
\State \quad\,\quad\,$v$ is referenced in $m.digests$
\State \quad\,\quad\,No referenced vertex is in the past of $v$
\State \quad\,\quad\,\Return true
\State \quad\,\textbf{end if}
\State \textbf{end if}
\State \Return false
\EndProcedure 
\end{algorithmic}
\end{algorithm}

To be more precise, we call a graph $\acute{G}$ an extension of a Lamport graph $G$, by a vertex $v$, if and only if $\acute{G} - G = v$, e.g if $G$ and $\acute{G}$ differ by $v$, only. As the following proposition shows, any extension of a Lamport graph, is itself a Lamport graph.
\begin{proposition}[Lamport graph extensions]
Suppose that $G$ is a Lamport graph and $m$ a byte string with $\textsc{Integrity}(m,G)=true$. Then the extension $\acute{G}$ of $G$ by a vertex $v$ with $v.m=m$ is a Lamport graph.
\end{proposition}
\begin{proof} The \textsc{message\_integrity} function implies that there is no other vertex in $\acute{G}$ that is equivalent to $v$ and that all direct causes $v\to \acute{v}$ of $v$ are elements of $G$, hence of $\acute{G}$.
\end{proof}
Message integrity detects most faulty behavior. However there is a kind of fault, called a \textit{mutation}, that can not be ruled out in this way, because it is not strictly local and therefore undetectable from the outside of any Lamport graph. Such a mutation occurs, if a set of messages, all with the same id, properly reference one and the same previous message with that $id$ in a Lamport graph. The set then mutates the causal chain of messages with the same $id$ and these errors are mapped into the Lamport graph. They are the reasons for byzantine agreement to appear in the first place.

No message integrity check can rule this out, as such a failure occures only relative to other messages and those messages might arrive at different processes during different times. Moreover since we assume no signature scheme, every process can generate mutations for any $id$ that it knows\footnote{However the system might be designed in such a way that certain $id$'s a economically favored over others, for example if a reward is associate to it that can only be accessed by the original creator of that $id$.}. 
\begin{definition}[Mutation] Let $G$ be a Lamport graph. Then two vertices $v$ and $\acute{v}$ in $G$ are called a \textit{mutation} of a virtual process, if they have the same id and are spacelike, i.e neither $v\leq \acute{v}$ nor $\acute{v} \leq v$ holds.
\end{definition}
\begin{remark}Mutations like this are called forks in the hashgraph consensus paper \cite{BL}, which describes the situation quite nicely. However we stick to the original term as defined by the actual providers Moser \& Melliar-Smith to properly honor their contribution as it should be.
\end{remark}
The possibility of mutations is the true reason, why total order algorithms from the Moser \& Melliar-Smith family need byzantine fault tolerance. Messages are considered as votes from virtual processes and mutations mimic byzantine behavior in actual voting systems, where an actor might deliver different votes to different processes. This is exactly the situation, that byzantine fault tolerant protocols deal with.

\subsection{Member discovery gossip}
We view the system as a dynamic, directed graph, where vertices are processes and an edge indicates the current ability of a process to send a message to another process. We follow \cite{ABFG} in their notation and write $\Pi(t)$ for this graph, as it would appear to an omnipotent outside observer. However no process must know the entire system and each $j\in \Pi(t)$ might have a partial view $\Pi_j(t)$ only. By definition, process $j$ is then able to send data to any member $k$ of its local view, but to no other participant. Our system therefore meets the criteria of a proper, unstructured Peer-to-Peer network. 

Assuming a solution to the bootstrapping problem, every honest process $j$, knows a partial view, strictly larger then itself. The first gossip protocol is then a classic \textit{process discovery gossip}. 

It consists of a standard push\&pull gossip, which means, that any honest process will choose another process periodically (but asynchronous, i.e. clock ticks are entirely local) at random and sends it a list of processes that it thinks are currently participating in the protocol and a list of processes, that it thinks have (temporally) left the system. In addition it will choose a random process, to ask it for a list of participating and leaving peers. In turn, if a honest process receives such a request, it sends a list of members that it thinks are currently participating in the network and a list of leaving processes in return. Algorithm (\ref{alg:discovery-gossip}) gives an example way to realize this protocol.

\begin{algorithm}
\caption{Process discovery}
\label{alg:discovery-gossip}
\scriptsize
\begin{algorithmic}[1]
\Statex \textbf{run the following two loops in parallel forever} 
\Statex
\Loop { }discovery push\&pull
	\State wait for Poisson clock tick
	\State send subset of $\Pi_j$ to random process $k\in\Pi_j$
	\State send discovery requests to random process $k'\in\Pi_j$
\EndLoop
\Statex
\Loop
	\State wait for data package
	\If{data is a set of processes}
		\State update $\Pi_j$
	\ElsIf{data is process discovery requests}
		\State respond with subset of $\Pi_j$
	\EndIf
\EndLoop
\end{algorithmic}
\end{algorithm}

As this goes on forever each process $j$ will have an ever changing partial view $\Pi_j(t)$ into the system and it is free to restrict the amount of neighbors $|\Pi_j(t)|$ it knows, to not store to much data. No stop argument is involved and the network load must be regulated by the participating processes them self.

Depending on the actual churn, the frequency of the communication, that is the rate of the local clock ticks might be rather low, for example in the range of minutes. The purpose of this protocol is just to keep $\Pi_j(t)$ up to date, which enables a process to send and receive data from other processes. 
\begin{remark}
A system engineer might incorporate additional stragegies to make communication etween honest processes more likely. However we leave this question open for further development.
\end{remark}

\subsection{Message gossip} 
Assuming that a process has a partial view $\Pi_j(t)$ into the network that is not completely wrong, it participates in the message gossip, which is the second asynchronous push\&pull gossip. Its purpose is to distribute messages through the current population $\Pi(t)$.
\begin{algorithm}
\caption{Message gossip}
\label{alg:message-gossip}
\scriptsize
\begin{algorithmic}[1]
\Statex \textbf{run the following loops in parallel forever} 
\Statex
\Loop { }send push\&pull
	\Comment{On many threads}
	\State wait for Poisson clock tick
	\State send $S\subset \{v.m\;|\; v\in G \,\wedge \, v.total\_position=\bot\}$ 
	\Statex \quad\, to random process $k\in\Pi_j$
	\State send request for missing messages to random 
	\Statex \quad\, process $k'\in\Pi_j$
\EndLoop
\Statex
\Loop { }receive
	\State wait for data package $m$
	\If {\textsc{message\_integrity}$(m,G)$}
		\State expand $G$ with vertex $v$, such that $v.m=m$
	\ElsIf{data is message requests}
		\State respond with appropriate set of messages
	\EndIf
\EndLoop
\end{algorithmic}
\end{algorithm}

Messages are retransmitted via push gossip, only if they don't have a total order yet. This is the 'stop' criterion, required by high frequency push gossip protocols in general. Already ordered messages are pushed only as a response to a pull request. 

However, despite such a stop criteria, message gossip never really stops, even if the production of new messages comes to a hold. This happens because without new messages, some of the previous ones might not achieve an order and are therefore retransmitted forever. For the system to be live we therefore have to make the assumption that new messages appear forever.

\section{Total Order}Crisis extends the timelike causality between messages into a probabilistically converging total order, that enables comparison of spacelike messages in an invariant way.

Convergence happens as long as the network is able to estimate the overall amount of voting weight per time and the majority of processes behind that weight are interested in a stable order. We call this estimation a \textit{difficulty oracle}, because it might behave very much like Bitcoins difficulty function in certain implementations. Fortunately, proper behavior is incentivizeable and deviation can be punished. The system therefore utilizes economical interest to achieve convergence. 

Total order is then generated in four basic steps: The Lamport graph is divided into rounds and each round is tested for the occurrence of a so called safe voting pattern. Every time such a pattern appears, a next step in Micali's player replaceable agreement protocol $BA^*$ is executed to locally decide a \textit{virtual round leader} vertex. 

Under partitioning, this leader might not be unique and a selection process similar to Bitcoins longest chain rule is applied such that all Lamport graphs eventually converge on the same round leader. In any case, the past of these round leader vertices is ordered concurrently to the rest of the system,  using some kind of topological sorting, like Kahn's algorithm in combination with the voting weight to decide spacelike vertices. As the virtual round leader converges to a fixed value, so does the order.

\subsection{Votes}As pioneered by Moser \& Melliar-Smith \cite{MM}, total order is achieved, if vertices vote on other vertices in some kind of virtual byzantine agreement process. Therefore, each vertex $v$ has a field $v.vote$, where the entry $v.vote(r)=(l,b)$ describes $v$'s vote $(l,b)$ on some message $l\in \textsc{message}$, together with a possibly undecided binary value $b\in\{\bot,0,1\}$ in a so called round $r$.

\subsection{Virtual communication} 
To appreciate the idea of virtual voting and to see how algorithm (\ref{alg:virtual-leader}) works, we need to understand the information flow between so called \textit{virtual processes} inside any given Lamport graph. In fact some virtual process $id$ is able to pull information from another virtual process $id'$, if and only if there are appropriate vertices $v$ and $\acute{v}$, such that $v.id=id$, $\acute{v}.id=id'$ and $\acute{v}$ is in the past of $v$. In that case, we say that virtual process $id$ received votes $\acute{v}.vote$ from virtual process $id'$ and that there is a communication channel from $v$ to $\acute{v}$.

Simulated communication channels like this are of course tamper proof and invariant among all Lamport graphs, due to the invariance of the past theorem (\ref{thm:invariant-past}). However byzantine behavior might still appear in the form of mutations and strategic, non random, message distribution.

Strategic message dissemination occurs, because any real world process is able to deviate from random gossip and send certain messages to certain peers only. In addition carefully mutated vertices might show different votes from the same virtual process, because any message creator is relatively free in choosing the past of any message. The overall effect is a virtual voting equivalent to the well known phenomena of byzantine actors sending different votes to different processes. A situation well suited for byzantine agreement protocols.

Although we have to accept such a behavior to some degree, it is nevertheless possible to prohibit strategic message distribution from any bounded adversary, who is able to manipulate an overall amount of voting weight $k$ only. 

One way to achieve this is by sending virtual votes through vertex disjoint path, only if the combined weight of their leightest vertices is greater then $k$. Such a strategy would be a weighted virtual interpretation of the message dissemination algorithm from \cite{ABFG}. 

However counting disjoint paths is computationally expensive and not really necessary in our setting, because virtual communication channels are already tamper proof from the inside. All we need is some insurance, that information flows through enough vertices from different real world processes. Such a requirement makes proper message distribution likely and counteracts any partition tendency to some degree. In a addition, it can be measured efficiently by just counting the overall weight in all path between two vertices.
\begin{definition}[$k$-reachability]
\label{def:k-reachability}
Let $k$ be a positive number, $G$ a Lamport graph and $v,\acute{v}\in G$ two vertices. Then $\acute{v}$ is said to be $k$-reachable from $v$, if the overall weight of all vertices in all path from $v$ to $\acute{v}$ is greater then $k$. In that case we write $\acute{v}\leq_{k} v$.
\end{definition}
If we interpret a Lamport graph such that a \textit{byzantine resistant} virtual communication channel exists from $\acute{v}$ to $v$ only if $\acute{v}\leq_k v$, we ensure that $k$-bounded collaborations can not influence virtual communication channels by strategic message distribution.

\subsection{Virtual synchronism} Lamport graphs represent a timelike order between vertices, that we interpret as virtual communication channels. Going one step further, we can forget about the outside world altogether and just think from the inside of a Lamport graph to define a virtual clock tick as a transition from one vertex to another. 

This simple idea allows for internal synchronism, that enables us to execute strongly synchronous agreement protocols like Feldman \& Micali's algorithm $BA^*$ \cite{SM} virtually, but without any compromise in external asynchronism. 
\begin{remark}
Again this insight was already present in the work of Moser \& Melliar-Smith \cite{MM} under the term 'stage'. In fact it appears quite natural from the perspective of the well known Lamport clocks.
\end{remark}

Any byzantine resistant protocol is based on the assumption that the amount of faulty behavior does not exceed a certain fraction of the overall voting weight and Crisis
utilizes such a threshold too. However in contrast to most approaches, consistency does not depend on it, e.g. the order does not fork, even if the bound gets broken from time to time. Crisis therefore favors consistency over availability in such a scenario. 

In any case, we need a way to approximate the voting weight, that is generated in a round and we must assume, that from time to time, not more then $1/3$ of this weight is faulty.  

We call such an approximation a \textit{difficulty oracle}, because it might behave very much like Bitcoins difficulty function in certain Proof-Of-Work based incarnations. In any case, it is considered as an external parameter and different choices might lead to different behavior.
\begin{definition}[Difficulty oracle] Suppose that the tuple $(\mathbb{W},w,c_{min},\oplus)$ is a weight system of the protocol. Then the function
\begin{equation}
d: \mathbb{N} \to \mathbb{W}
\end{equation}
that maps natural numbers onto weights, is called a \textbf{difficulty oracle} w.r.t. the weight system and the value $d_r:=d(r)$ is called the \textbf{round $r$ difficulty} of the system.
\end{definition}
\begin{example}
The most simple example would be to just use a fixed constant that does not change over time as the systems difficulty oracle. This however might be way to simple for certain choices of weight systems, as we know from protocols like Bitcoin, that the overall voting weight per time (hash power for that matter) might fluctuate considerably. The overall goal is to compute a difficulty oracle such that equation (\ref{eq:difficulty-bound}) holds approximately.
\end{example}
\begin{example}
A more flexible difficulty oracle would hardcode its value for the first few rounds and then base the computation on the overall voting weight in the past of converged virtual round leader vertices later on. That past is invariant and therefore every process would compute exactly the same function. 
\end{example}

With such a difficulty oracle at hand, we can look at algorithm $BA^*$ as explained in \cite{SM}, to see that it executes a potentially unbounded amount of synchronous rounds, each of which starts with a communication step, where any actor receives votes broadcast by the actors of the previous round. Our goal in this section is therefore to simulate that behavior, using the idea of internal time in combination with our byzantine safe communication channels (\ref{def:k-reachability}). Algorithm (\ref{alg:sync-clocks}) gives the details.
\begin{algorithm}
\caption{Virtual synchronous rounds}
\label{alg:sync-clocks}
\scriptsize
\begin{algorithmic}[1]
\Require 
\Statex connectivity $k$
\Statex difficulty oracle $d$
\Statex
\Procedure{round}{vertex:v, lamport\_graph:G}
\State $N_v\leftarrow \{\acute{v}\in G \;|\; v\to \acute{v}\}$
\label{alg.virtual-synchronism-line-start}
\State $r\leftarrow max(\{\acute{v}.round \;|\; \acute{v}\in N_v \}\cup\{0\})$
\If {there is a $\acute{v}\in N_v$ with $\acute{v}.is\_last$ and $\acute{v}.round = r$}
\label{alg.virtual-synchronism-line-first-if}
\State $v.round \leftarrow r+1$
\label{alg.virtual-synchronism-line-v-round-r+1}
\Else
\State $v.round \leftarrow r$
\label{alg.virtual-synchronism-line-v-round-r}
\EndIf 
\State $S_v\leftarrow \{\acute{v}\in G\;|\;\acute{v}.round=v.round-1, 
	     \acute{v}.is\_last,  \acute{v}\leq_{k} v\}$
\If{$w(S_v)> 3\cdot d_r$}
\label{alg.virtual-synchronism-line-second-if}	
\State $v.last\leftarrow true$
\Else
\State $v.last\leftarrow (r=0)$
\label{alg.virtual-synchronism-line-v-last-false}
\EndIf
\EndProcedure
\Statex
\Statex The procedure assumes a previous execution on all vertices in the past of $v$, but it can be called concurrently on spacelike vertices. 
\end{algorithmic}
\end{algorithm}

The algorithm computes so called \textit{round numbers} and the $is\_last$ property of any vertex. The round number of a vertex is computed by first taking the largest round of all direct causes as its current estimation. If the vertex is a direct effect of a current round vertex with the $is\_last$ property, a new round begins and the vertex is a first vertex in that new round. If the vertex has enough last vertices of the previous round in its past and it is $k$-reachable from all of them, the vertex becomes a last vertex in its own round. 

Last vertices are interpreted, as sending and receiving votes through byzantine resistant virtual communication channels to and from last vertices of consecutive rounds. This way, last vertices model the behavior of actors sending votes to other actors, whenever a round transition happens in algorithm $BA^*$.

\begin{remark}The appearance of new rounds can not be guaranteed, even if we assume new messages to arrive forever. This is because the required high interconnectivity between messages must not happen. Extreme situations are thinkable, where no message references any other message and the Lamport graph is totally disconnected. Then, of course, no interconnectivity occurs and all message have a round number of zero forever.
Any actual incarnation therefore requires proper incentivation to encourage the appearance of new rounds. This is possible, for example, if only vertices with $v.is\_last=true$ are incentivised in one way or another by the systems incentivation function.
\end{remark}

Now, to understand our concept of virtual rounds a bit better, we proof a series of statements, that basically show that the round number and the $is\_last$ property are well defined and behave as expected. We start by showing that both properties do not depend on the actual Lamport graph, but are the same for equivalent vertices.
\begin{proposition}[Round invariance] Let $v$ and $\acute{v}$ be two equivalent vertices in Lamport graphs $G$ and $\acute{G}$ respectively. Then $v.round=\acute{v}.round$ and $v.is\_last=\acute{v}.is\_last$.
\end{proposition}
\begin{proof}Both, the round number and the $is\_last$ property depend on certain sets of vertices in the past of a vertex, only. But since $v$ and $\acute{v}$ are equivalent, they have equivalent pasts $G_v$ and $\acute{G}_{\acute{v}}$, due to the invariance of the past theorem (\ref{thm:invariant-past}). We can therefore proof the statement by strong induction on the number of vertices, both in $G_v$ and $\acute{G}_{\acute{v}}$. 

For the base case assume that $G_v$ contains $v$ only. In that case $\acute{G}_{\acute{v}}$ contains $\acute{v}$ only and both $N_v$ and $N_{\acute{v}}$, are empty. Then $v.round=0$ and $\acute{v}.round=0$, since algorithm (\ref{alg:sync-clocks}) executes line (\ref{alg.virtual-synchronism-line-v-round-r}) in both cases. Moreover, $S_v$ and $S_{\acute{v}}$ are empty which implies $v.is\_last=true$ and $\acute{v}.is\_last=true$, since algorithm (\ref{alg:sync-clocks}) executes line (\ref{alg.virtual-synchronism-line-v-last-false}) in both cases and $r=0$.

For the induction step, assume that $G_v$ and $\acute{G}_{\acute{v}}$ are given and that $x.round=\acute{x}.round$ as well as $x.is\_last=\acute{x}.is\_last$ holds for all equivalent verices $x$ and $\acute{x}$ in all Lamport graphs, with $|G_x|< |G_v|$ as well as $|G_{\acute{x}}|< |G_v|$. 

Then there is exactly one $\acute{x}\in N_{\acute{v}}$ for every $x\in N_v$ and $x.round=\acute{x}.round$ as well as $x.is\_last=\acute{x}.is\_last$, since $N_v$ and $N_{\acute{v}}$ are equivalent by the invariance of the past theorem (\ref{thm:invariant-past}) and $|G_x|< |G_v|$ as well as $|G_{\acute{x}}|< |G_v|$. This however implies $v.round=\acute{v}.round$, because algorithm (\ref{alg:sync-clocks}) computes the same value $r$ both for $v$ and $\acute{v}$ and decides the same branch in line (\ref{alg.virtual-synchronism-line-first-if}).

A similar reasoning shows $x.round=\acute{x}.round$ as well as $x.is\_last=\acute{x}.is\_last$ for all $x\in S_v$ and $\acute{x}\in S_{\acute{v}}$ and that $w(S_v)=w(S_{\acute{v}})$ holds, since equivalent vertices have equal weight. Hence both executions of algorithm (\ref{alg:sync-clocks}) decide the same branch in line (\ref{alg.virtual-synchronism-line-second-if}) and the proposition holds on $G_v$ and $\acute{G}_{\acute{v}}$, which proof the proposition in any case by strong induction.
\end{proof}

Round numbers are compatible with causality, in the sense that the round number of a future vertex is never smaller then the round number of any vertex in its past. Round numbers are therefore an important first step in any attempt to totally order a Lamport graph. The following proposition gives the details.
\begin{proposition}
\label{thm:monotone-round-number}
Let $v$ and $\acute{v}$ be two vertices in a Lamport graph $G$, such that $\acute{v}\leq v$ holds. Then $\acute{v}.round \leq v.round$.
\end{proposition}
\begin{proof}To see this, first assume $v\to \acute{v}$. Then $\acute{v}\in N_{v}$ and algorithm (\ref{alg:sync-clocks}) computes $v.round\geq \acute{v}.round$ and the statement holds. The general situation then follows by repeated execution of (\ref{alg:sync-clocks}) on each vertex in the causal chain $v = v_1 \to v_2 \ \cdots v_{n-1} \to v_n = \acute{v}$.
\end{proof}
A vertex has the $is\_last$ property, if and only if it is indeed a last vertex in a given round, i.e. every vertex in its future has a higher round number. We can therefore interpret these vertices as the end of a step in virtual $BA^*$ and as the exact point in internal time, where a virtual process sends its vote to members of the next step. This serves as the basis for our virtual adaptation of algorithm $BA^*$.
\begin{proposition}[Last vertices of a round]Let $v$ be a vertex in a Lamport graph, with $v.is\_last=true$. Then every vertex in the future of $v$ has a round number, strictly larger then $v.round$.
\end{proposition} 
\begin{proof}Let $\acute{v}$ be a vertex in the future of $v$. Then, there is a path $\acute{v}\to v_1 \to \cdots \to v_k \to v$ in any Lamport graph, that contains $\acute{v}$ (and therefore $v$) and $\acute{v}.round \geq v_k.round$ follows from proposition (\ref{thm:monotone-round-number}).

However, since $v\in N_{v_k}$ and $v.is\_last = true$, either $max\{\tilde{v}.round \;|\; \tilde{v}\in N_v \}> v.round$, or algorithm (\ref{alg:sync-clocks}) executes line (\ref{alg.virtual-synchronism-line-v-round-r+1}). In any case, the round number of $v_k$ is strictly larger then the round number of $v$ and we get $\acute{v}.round \geq v_k.round > v.round$. 
\end{proof} 
If we consider vertices of a given round to receive votes from vertices of a previous round, we have to be sure, that those previous round vertices are indeed in the past of any current round vertex. The following proposition shows that this is indeed the case.
\begin{proposition}
\label{thm:past-has previous-rounds}
Let $G$ be a Lamport graph and $v$ a vertex with a positive round number $v.round>0$ in $G$. Then $v$ has at least one last round $s$ vertex in its past for all round numbers $s<r$. 
\end{proposition} 
\begin{proof}
We show the proposition for $s=r-1$. The general case then follows by recursion, since $G$ is finite.

To see the statement, observe that for a vertex to be in round $r$, $r$ must either be the largest round number of its direct causes, or it must have a direct cause of round number $r-1$ that is a last vertex of that round. 

The second case is immediate. For the first case the argument can be repeated with any round $r$ direct cause. Since the graph is directed, acyclic and finite and any sink vertex has round $0$, there must eventually be a round $r$ vertex, that has no round $r$ direct causes. 
\end{proof}
\subsection{Difficulty bounds} 
\label{sec:difficulty-bounds}
Both the voting weight and the number of virtual processes is potentially unbounded in any given round. It is true that we can approximately limit the amount of faulty behavior at every moment by proper incentivation and punishment, but given enough time, byzantine behavior accumulates in the graph. In fact everyone can add arbitrary amounts of vertices with arbitrary large weights into any round, provided that sufficiently many new rounds appeared ever since.
 
For example a system with a Bitcoin-style Proof-Of-Work voting weight, might observe the occasional occurrence of something like a hash-bomb, i.e. a super heavy message that suddenly appears, but references messages way back in the past only. Such a 'bomb' is able to break all global byzantine bound assumptions in any round and it certainly exists if some motivated process puts all its hashing power for weeks, or even years into the generation of just one single message.

Another extreme example would be some kind of Internet-meme like phenomena, where suddenly large amounts of small to medium size messages occur in very old rounds for no apparent reason. In particular anybody can generate new messages in round zero easily, by not referencing other messages at all. 

We might call fringe cases like this \textit{time travel attacks}. The underlying reason is, that byzantine behavior is unbounded altogether, despite the fact that we can assume it to be approximately bounded at any given moment in time by our difficulty oracle. 

According to Brewers CAP-theorem, behavior like this is unavoidable in any open and asynchronous system, because partition happens in unstructured systems without any governance, stake-, or member-lists in one way or another. It sharply distinguishes our situation from more traditional approaches like hashgraph \cite{BL}, parsec \cite{CKHMS}, or blockmania \cite{DH} and puts our algorithm much more closely to Nakamoto's consensus.

Thats being said, unbounded byzantine behavior never happens in the past of any vertex,  because that past is fixed forever, due to the invariance of the past theorem (\ref{thm:invariant-past}). We can therefore counteract such an attack locally, by carefully computing all relevant properties relative to the perspective of a vertex in a consecutive round only. The price to pay is globality, because agreement is achieved locally only.
 
Now, ideally, that is in an imaginative system without partitions, a system engineer would design the difficulty oracle such that the overall voting weight $w^G_r$ of last messages in round $r$ of Lamport graph $G$ would always be in the range $3\cdot d_r < w^G_r \leq 6\cdot d_r$. This would guarantee any local round leader to be the global round leader and the order would be strictly convergent, not just probabilistically.

However, time travel attacks, forking and partitions are something to consider and because of that, the overall voting weight of a round is undefined, must not converge and varies between different Lamport graphs. The difficulty oracle can therefore be designed in such a way that an overall weight $>3\cdot d_r$ eventually happens frequently, but an upper bound estimation is impossible in general.  

On the other hand, it is still rational to assume that the voting weight \textit{per time} is in a certain range, at least approximately. Fortunately, this is enough to compute a theoretical upper bound on the overall amount of voting weight that might occur in any Lamport graph $G$. This bound can then be used to guarantee probabilistic convergence of the total order.

To see that, let $t$ be an external time parameter and $\Pi(t)$ the system at time $t$. Then the maximal round number $r_t$ at time $t$ is the maximum of all round numbers in all Lamport graphs of the system $\Pi(t)$ as it would appear to an omnipotent external observer\footnote{Of course this number is entirely theoretical as no participant can actually know it.}. 

We can use this number to give an upper bound on the amount of voting weight that occurs in the system. If $G_{max}(t)$ is the largest Lamport graph that exists in the system at time $t$ and if $w^G_s$ is the overall voting weight of all last vertices in some Lamport graph $G$ that has a round number $s$, then we assume our difficulty oracle to be designed such that
\begin{equation}
\label{eq:difficulty-bound}
\lim_{|G| \to |G_{max}(t)|} \sum_{s=0}^{r_t} \frac{w^G_s}{d_s}  \leq 6
\end{equation} 
holds \textit{approximately} for all external time parameters $t$. Of course this number is theoretical, as no actual process can compute it, because no process knows $r_t$, or $G_{max}(t)$.

Basically, this inequality expresses the idea that the difficulty oracle is designed such that the amount of voting weight per time, is limited and no more then $6 d_s$ weight can be produced in any round on average. However, it is flexible enough to allow every process to append generated voting weight into any round that currently exists.
 
\subsection{Virtual process sortition}
In \cite{ABFG}, Alchieri et al. looked at byzantine agreement in systems with unknown participants (BFT-CUP) and gave sufficient conditions to solve it. Their reasoning is solid, but they didn't consider player replaceable protocols. 

However with player replaceability in mind, the situation changes, because every step in the protocol is executable in an entirely different set of processes. This implies that new solutions might appear and indeed a family of such solutions was found by Chen \& Micali in Algorand \cite{CM}, where consensus in open systems becomes more or less a problem of synchrony and quorum selection. The latter of which can be nicely solved by cryptographic sortition.

Nevertheless, the present situation is somewhat orthogonal to Algorand, as we can simulate synchronism easily, but cryptographic sortition might not work in our virtual setup. We therefore face the problem of how to decide, which virtual processes should execute a step in the protocol. Moreover, as our system is open and asynchronous, an unbounded amount of virtual processes might appear in any round. 

Fortunately we can put things into perspective and consider the past of a vertex only, which fixes the problem of unbounded vertices, relatively speaking. However there might still be too much entropy in the system and we need a way to deterministically compute a subset of virtual processes that is somewhat favorable in the execution of a next step in the agreement protocol. 

We call such a mechanism a \textit{quorum selector} and consider it as another important parameter in any actual incarnation. Like the voting weight, different quorum selector functions might lead to very different long term behavior and the author believes that it is currently impossible to decide which one performs best under any given circumstances. 

In any case, quorum selector functions decide virtual processes, not vertices. We therefore need a way to go from vertices to virtual processes first. This however is efficiently done, by deriving another graph from any Lamport graph, that projects vertices of equal $id$'s together. The following two definitions make the idea precise.
\begin{definition}[Relative subgraph of a round]
Let $s$ and $r$ be two round numbers with $s<r$, $G$ a Lamport graph, $v$ a round $r$ vertex in $G$ and $V_v^s$ the set of all round $s$ vertices in the past of $v$. Then the subgraph $G_v^s:= (V_v^s, A_v^s)$ of $G$, with $(x,y)\in A_v^s$, if and only if $x,y\in V_v^s$ and $x\to y$, is called $v$'s \textbf{round $s$ past} in $G$. 
\end{definition}
Now, the transition from vertices to virtual processes is done, by collapsing all vertices with the same id into some kind of new meta-vertex in the so called quotient graph. The latter of which is nothing but a quotient object in the category of graphs.
\begin{definition}[Knowledge graph]Let $s$ and $r$ be two round numbers with $s<r$, $G$ a Lamport graph, $v$ a last message in round $r$ and $G_v^s$ the round $s$ past of $v$ in $G$. Then the quotient graph $\Pi_v^s:= G_v^s\backslash \simeq_{id}$ defined by the equivalence relation $x \simeq_{id} y$, if and only if $x,y\in G_v^s$ and $x.id=y.id$, is called $v$'s \textbf{round $s$ knowledge graph}.

We write $id$ for an equivalence class vertex $\{\acute{v}\in G_v^s\;|\; \acute{v}.id=id\}\in \Pi_v^s$ and call it a round $s$ \textbf{virtual process}, from the perspective of $v$. 
\end{definition}
Given any Lamport graph, our definition of knowledge graphs is efficiently computable and can be stored with little additional overhead. It is directed, but in general not acyclic anymore and it resembles a virtual version of the knowledge connectivity graph from \cite{ABFG}. 

To understand the meaning of this graph, consider that a virtual process $id\in \Pi_v^s$ has a directed edge to another virtual process $id'\in \Pi_v^s$, if and only if there is a vertex $\tilde{v}$ with $\tilde{v}.id=id$ and a vertex $\acute{v}$ with $\acute{v}.id=id'$ in $G_v^s$, such that $\tilde{v}\to\acute{v}$. Hence any edge represents the knowledge a virtual process has about the existence of another virtual process relative to a given round.

The following two propositions show that knowledge graphs are indeed well defined and invariant among different Lamport graphs.
\begin{proposition}[Existence]Let $s$ and $r$ be two round numbers with $s<r$, $G$ a Lamport graph and $v$ a last message in round $r$. Then the round $s$ knowledge graph $\Pi_v^s$ is well defined, directed and not empty.
\end{proposition}
\begin{proof}
Since $v$ is a last vertex in a round $r>s$, $v$ must have round $s$ vertices in its past due to proposition (\ref{thm:past-has previous-rounds}). This however implies, that $G_v^s$ is not empty as a directed graph. In addition $\simeq_{id}$ is an equivalence relation on the vertex set of $G_v^s$, which implies that the quotient is a well defined, directed and not empty, by the general properties of quotient objects in the category of graphs. 
\end{proof} 
We call two knowledge graphs $\Pi_{v}^s$ and $\Pi_{\acute{v}}^s$ equivalent, if their reference vertices $v$ and $\acute{v}$ are equivalent. As the following proposition shows, equivalent knowledge graphs are isomorphic and their elements consist of equivalent vertices only. 
\begin{proposition}[Invariance of knowledge graphs] Let $s$ and $r$ be two round numbers with $s<r$ and $v$ as well as $\acute{v}$ two equivalent round $r$ vertices in Lamport graphs $G$ and $\acute{G}$, respectively. Then the knowledge graph $\Pi_v^s$ of $v$ is isomorphic to the knowledge graph $\acute{\Pi}_{\acute{v}}^s$ and the elements in each equivalence class $id\in \Pi_v^s$ are in one-to-one correspondence with equivalent elements in $id'\in \acute{\Pi}_{\acute{v}}^s$.
\end{proposition}
\begin{proof}

The invariance of the past theorem (\ref{thm:invariant-past}) implies, that $G_v^s$ and $G_{\acute{v}}^s$ are isomorphic and vertices with equal $id$'s are in one-to-one correspondence. Hence their quotients under the $\simeq_{id}$ relation, are isomorphic. Moreover, each equivalence class $id\in \Pi_v^s$ consist of vertices from $G_v^s$ that have the same id. However due to invariance of the past, these are in one-to-one correspondence with vertices in $\acute{G}_{\acute{v}}$ that project onto the appropriate id in $\acute{\Pi}_{\acute{v}}^s$.
\end{proof}
Now, given any knowledge graph, a quorum selector is nothing but a way to chose a subset of virtual processes from that graph. The members are then interpreted as to send and receive votes through their last vertices. 
\begin{definition}[Quorum selector] Let $s$ and $r$ be two round numbers with $s<r$, $v$ a last round $r$ vertex in a Lamport graph $G$ and $\Pi_v^s$ the round $s$ knowledge graph of $v$. Then a \textbf{quorum selector} \textsc{quorum} deterministically chooses a subset $Q_v^s\subset\Pi_v^s$, called $v$'s round $s$ quorum, such that $Q_v^s$ and $Q_{\acute{v}}^s$ are equivalent for equivalent graphs $\Pi^s_v$ and $\acute{\Pi}^s_{\acute{v}}$.
\end{definition}
Quorum selection serves as a kind of filter, to reduce the overall byzantine noise, that might appear in the voting process of fully open systems. Its purpose is to make the appearance of a so called \textit{safe voting pattern} as defined in the next section, more likely. 

\begin{example}[Highest voting weight quorum] Voting weight of vertices can be combined into voting weight of appropriate equivalence classes in $\Pi_v^s$, if we define $w(id):= \bigoplus_{v\in id} w(v)$ for any $id\in \Pi_v^s$. This is invariant among equivalent knowledge graphs $\Pi_v^s$ and $\Pi_{\acute{v}}^s$ and low weight mutations do not change that value much.

A quorum selector function is then given by first choosing the weakly connected component of $\Pi_v^s$, that has the highest combined voting weight and then by ordering all virtual processes in that component according to their individual weight. After that the quorum selector might takes the heaviest $n$ vertices from it, where $n$ is a suitable constant, that makes the appearance of enough last vertices with an overall voting weight strictly larger then $3d_s$ probable. 

The reasoning here is, that by restricting to a weakly connected  component, faulty behavior based on graph partition is reduced. Moreover different vertices will compute the same quorums, as it is unlikely that the voting weight will fluctuate that much, seen from the perspective of different vertices. Moreover, mutations will effect the votes of these sets the least, simply because the voting power of very heavy vertices is less affected by lightweight mutations. 
\end{example}

\subsection{Safe voting pattern} 
With a quorum selector function at hand, we can now look at the last vertices of all quorum members in a given round and see if they qualify as proper voting sets. 

Similar to any other byzantine agreement protocol, our virtual leader election (\ref{alg:virtual-leader}) is based on the assumption that the amount of faulty behavior is bounded and does not exceed a certain amount of the overall voting weight. If this holds true voting takes place, if not voting stalls until the situation eventually resolves.

The purpose of a \textit{safe voting pattern} is therefore to make sure, that voting takes place in those rounds only, that have appropriately bounded byzantine behavior. As described in section (\ref{sec:difficulty-bounds}), the overall amount of faulty behavior is necessarily unbounded in any round, as the system is open and fully asynchronous. However it is always bounded relative to the past of any vertex, simply because that past is frozen and does not change ever again, due to the invariance of the past theorem (\ref{thm:invariant-past}). 

This leads naturally to our definition of safe voting patterns, but before we derive the details, we need to specify the concept of a voting set first.
\begin{definition}[Voting sets] Let $k\in \mathbb{R}^+$ be a positive number, $r$ and $s$ two round numbers with $s<r$ and $v$ a last round $r$ vertex in a Lamport graph $G$. Then the set
\begin{multline*}
S_v(s,k):=\{x \;| \; x.id \in Q(v,s) \wedge  x \leq_{(r-s)k} v   \\
\wedge\; x.round=s \, \wedge \,  x.is\_last=true \}
\end{multline*}
is called $v.id$'s round $s$ \textbf{voting set} and $v.id$ is said to receive voting weight from the members of $Q(v,s)$ through $S_v(s,k)$. In addition, if $t$ is another round number, with $t<s$, $l\in\textsc{message}$ a message and $b\in\{\bot, 0,1\}$ a possibly undecided binary value, then 
\begin{multline*}
w(S_v(s,k),t, (l,b)):=   \\
w(\{x\in S_v(s,k)\;|\; x.vote(t)=(l,b)\})
\end{multline*}
is called the overall voting weight for the round $t$ vote $(l,b)$ that $v.id$ receives from its voting set $S_v(s,k)$.

We moreover say that $v$ receives a \textbf{super majority} of voting weight for a round $t$ vote $(l,b)$ from its voting set, if $w(S_v(s,k),t, (l,b))> w(S_v(s,k))\ominus d_s$ and a \textbf{honest majority} of voting weight, if $w(S_v(s,k),t, (l,b))> d_s$, where $d_s$ is the difficulty oracle in round $s$.
\end{definition}
Voting sets are invariant among equivalent vertices in different Lamport graphs, due to the invariance of the past theorem and the same holds for voting weights w.r.t. any given vote. The following proposition proofs the first statement, however to proof the second one, we need to understand how voting weights are actually computed first. We will do this in the following section.
\begin{proposition}[Voting set invariance] Let $v$ and $\acute{v}$ be two equivalent vertices in Lamport graphs $G$ and $\acute{G}$ respectively. Then the voting sets
$S_v(s,k)$ and $S_{\acute{v}}(s,k)$ are equivalent, i.e. both sets are isomorphic and consists of equivalent vertices only.
\end{proposition}
\begin{proof}Since the quorum selector is assumed to be invariant w.r.t. to vertex equivalence, all defining properties are actually invariant, which in tuen implies the invariance of any voting set.
\end{proof}
Using our definition of voting sets, we are now able to compute a safe voting pattern in a round. Algorithm (\ref{alg:safe-voting-pattern}) gives the details and we assume that it is executed on any vertex after algorithm (\ref{alg:sync-clocks}) only.
\begin{algorithm}
\caption{Safe voting pattern}
\scriptsize
\label{alg:safe-voting-pattern}
\begin{algorithmic}[1]
\Require 
\Statex connectivity $k$
\Statex difficulty oracle $d$
\Statex
\Procedure{svp}{vertex:v, lamport\_graph:G} 
\State $v.svp\leftarrow \emptyset:\emptyset$
\Comment empty total order
\State \textbf{if} $v.is\_last$ \textbf{and}
\label{alg:safe-voting-pattern-first-if}
\Statex \quad\,\quad\, there is a $\acute{s}<v.round$ with
\Statex \quad\,\quad\,\quad\, $3d_{\acute{s}} < w(S_v(\acute{s},k))\leq 6d_{\acute{s}}$ \textbf{and}
\Statex \quad\,\quad\,\quad\, $x.svp=y.svp$ for all $x,y\in S_v(\acute{s},k)$ \textbf{and} 
\Statex \quad\,\quad\,\quad\, ($x.svp\neq \emptyset$ \textbf{or} $s=0$) \textbf{and}
\Statex \quad\,\quad\,\quad\, $|w(S_{x}(t,k),u,(l,\bot))\ominus w(S_{y}(t,k),u,(l,\bot))|\leq d_{t}$
\Statex \quad\,\quad\,\quad\, $|w(S_{x}(t,k),u,(\cdot,b))\ominus w(S_{y}(t,k),u,(\cdot,b))|< d_{t}$
\Statex \quad\,\quad\,\quad\, $t\leftarrow max(x.svp)$ for $x\in S_v(\acute{s},k)$
\Comment $max(\emptyset)=-\infty$
\Statex \quad\,\quad\,\quad\, $\forall x,y\in S_v(\acute{s},k)$, rounds $u\in x.svp\backslash\{t\}$, votes $(l,b)$ 

\Statex \quad\, \textbf{then}
\State \quad\,$s\leftarrow$ maximum of all such $\acute{s}$
\State \quad\,$v.svp\leftarrow x.svp \cup\{s\}: s\leq s$ and $t<s$ for all $t\in x.svp$
\State \textbf{end if}
\EndProcedure
\Statex
\Statex The procedure assumes a previous execution on all vertices in the past of $v$, but it can be called concurrently on spacelike vertices. 
\end{algorithmic}
\end{algorithm}

Given any vertex $v$, algorithm (\ref{alg:safe-voting-pattern}) computes the totally ordered set $v.svp$, which is used to index round numbers that have safe voting patterns in the past of $v$. In particular, a voting set $S_v(s,k)$ is said to be a safe voting pattern, if $s$ is the maximal round number, such that $S_v(s,k)$ has enough overall voting weight to execute a step in a byzantine agreement protocol, all members $x\in S_v(s,k)$ have equal total orders $x.svp$ and all safe voting patterns of all members do not differ too much in any of their votes on previous rounds.

In addition, algorithm (\ref{alg:safe-voting-pattern}) implies, that safe voting patterns are nested sequences, where the elements of one stage reference the elements of a previous stage and so on. The following proposition makes this precise.
\begin{proposition}
\label{thm:svp-of-svp-member}
Let $v$ be a vertex with $v.svp\neq \emptyset$, $r=max(v.svp)$ and let $S_v(r,k)$ be $v$'s safe voting pattern. Then $x.svp=v.svp\backslash\{r\}$ for all $x\in S_v(r,k)$.
\end{proposition}
\begin{proof}
If $x\in S_v(r,k)$, then algorithm (\ref{alg:safe-voting-pattern}) computes the set $v.svp$ as $x.svp\cup\{r\}$.
\end{proof}
To properly speak about the distance between two safe voting patterns it is moreover  advantageous to define a metric on any totally ordered set $v.svp$.
\begin{definition}[Svp distance]Let $v$ be a vertex with $v.svp\neq \emptyset$. Then the \textbf{svp distance} is the function
\begin{equation}
d_{v.svp}: v.svp \times v.svp \to \mathbb{R}
\end{equation}
where $d_{v.svp}(r,r)=0$ and $d_{v.svp}(s,r)$ is otherwise defined for any $s,r\in v.svp$ with $s\neq r$ as the number of different elements between $s$ and $r$ in the internal order plus one.
\end{definition}

\begin{remark}
Safe voting patterns are not guaranteed to exist in any round, for various reasons. One of which is that the voting weights might differ to much, due to too much mutations. It is therefore of importance for any system engineer to implement some way that makes safe voting pattern at least likely. Ideally exactly one safe voting pattern would appear in every round. The more the system deviates from this rule, the more rounds are needed to make progress in the total order generation.
\end{remark}
On the bright side we know, that safe voting patterns are byzantine fault detectors, because they accurately measure the amount of mutations of quorum members. This is good news, as any such fault detector can then be used to implement some invariant way of incentivation and punishment, which in turn can be used to make safe voting patterns attractive and economically favorable. Moreover, the folk theorems of repeated games suggest that such a system can be guided into all kinds of behaviors.
\subsection{Local leader election}
\label{sec:virtual-leader-election}
Any safe voting pattern provides an environment for the execution of another step in a player replaceable byzantine agreement protocol. The algorithm we use is an adaptation of Chen, Feldman \& Micali's protocol $BA^*$, to the setting of Moser \& Melliar-Smith's idea of virtual voting on causality graphs in a BFT-CUP environment.

Loosely speaking, a local round leader is nothing but a message, that defines an invariant set of vertices in any Lamport graph, the latter of which is then integrated into the total order, using some kind of topological sorting. Leader messages are computed in a byzantine agreement process, because we need to be sure, that all Lamport graphs of honest processes agree on them, at least locally, i.e in the causality cone of a safe voting pattern.

In any case, execution of the agreement protocol start with an initial round leader proposal, computed by a so called \textsc{initial\_vote} function.
\begin{definition}[Initial Vote]Let $2^{\textsc{vertex}}$ be the power set of our vertex type. Then an \textbf{initial vote} function is a map
\begin{equation}
\textsc{initial\_vote}: 2^{\textsc{vertex}} \to  \textsc{message}
\end{equation}
that deterministically chooses a message from any given set of vertices, such that the outcome is the same for equivalent vertex sets.
\end{definition}
Initial vote functions are a system parameter and different choices might lead to different long term behavior. Ideally, all members of a safe voting patter would always compute the same initial vote. In that case an actual virtual round leader $l\neq\oslash$ would be decided in just a few extra rounds. However due to mutations, different members might compute different initial votes. In that case, it is the task of the virtual leader election to agree on a message anyway.

Since it is almost never the case that all members of a safe voting pattern are in agreement on a leader right away, the next best thing is to have at least a super majority of voting weight for some message. Based on this insight, the following example might give a reasonable choice for an initial vote function, based on the voting weight of messages.
\begin{example}[Highest weight] A simple yet fast implementation of the \textsc{initial\_vote} function is given by choosing the underlaying message of the highest voting weight vertex. Since we assume that it is infeasible to have different vertices of equal weight, such a choice is practically deterministic and the outcome depends on the underlying message only. 
\end{example}
After initial votes are made, a byzantine agreement protocol is executed in a chain of safe voting pattern, that locally decides on a message. However as the system is open, asynchronous and the voting weight is eventually unbounded in any round, we can never rule out, that a different leader is decided for the same round in another partition of the system. 

Because of that, algorithm (\ref{alg:virtual-leader}) itself does not decide a global leader but adds any local decision to the set of all possible leader in a round. The result is a stream of candidate sets that we call the \textit{global leader stream} of a Lamport graph.
\begin{definition}[Leader Stream]Let $G$ be a Lamport graph and  $2^{(uint, \textsc{message})}$ be the power set of indexed vertices. Then the function 
$$
\textsc{leader}_G: \mathbb{N} \to Option\langle 2^{(uint,\textsc{message})}\rangle
$$
is called the \textbf{global leader stream} of the Lamport graph, the set $\textsc{leader}_G(r)$ is called the \textbf{candidate set} for the virtual round $r$ leader and some element $(s,l)\in \textsc{leader}_G(r)$ is a possible round $r$ leader message $l$, locally decided in round $s$.
\end{definition}

\begin{algorithm}
\caption{virtual leader elections}
\label{alg:virtual-leader}
\scriptsize
\begin{algorithmic}[1]
\If {$v.svp=\emptyset$} 
\State $\textsc{leader}_G(r)\leftarrow \textsc{nakamoto}(\textsc{leader}_G(r),\oslash,v.round)$
\label{alg:virtual-void-leader-if-svp-none}
\State \Return
\label{alg:virtual-leader-return-if-svp-none}
\EndIf 
\State $s\leftarrow max(v.svp)$
\State $S \leftarrow$ $v$'s safe voting pattern $S_v(s,k)$
\State $n\leftarrow w(S)$
\For{all $t\in v.svp$}
\State $\delta \leftarrow d_{v.svp}(s,t)$ 
\If{$\delta = 0$}
\Comment{Initial leader proposal}
\State $v.vote(t) \leftarrow (\textsc{initial\_vote}(S),\bot)$
\Else
\State $l\leftarrow $ message with highest round $t$ voting weight in $S$
\label{alg:virtual-leader-l-for-positive-delta}
\If{$\delta = 1$}
\Comment{Leader presorting}
\If {$w(S,t,(l,\bot)) > n-d_{s}$}
\State $v.vote(t) \leftarrow (l,\bot)$
\Else 
\State $v.vote(t) \leftarrow (\oslash,\bot)$
\EndIf
\ElsIf {$\delta = 2$}
\Comment{$BBA^*$ initialization}
\If {$l\neq\oslash$ \textbf{and} $w(S,t,(l,\bot)) > n-d_{s}$}
\State $v.vote(t) \leftarrow (l,0)$
\ElsIf {$l\neq\oslash$ \textbf{and} $w(S,t,(l,\bot)) > d_{s}$}
\State $v.vote(t) \leftarrow (l,1)$
\Else 
\State $v.vote(t) \leftarrow (\oslash,1)$
\EndIf
\Else
\If {$\delta \mod 3 =0$}
\Comment{Coin fixed to $0$}
\If {$w(S,t,(l,0)) > n-d_s$}
\State $v.vote(t) \leftarrow (l,0)$
\label{alg:virtual-leader-l-c-f-t-z-super-majo}
\If {$w(S,t,(l,0)) = n$}
\label{alg:virtual-leader-agreement-on-l}
\State $\textsc{long\_chain}(\textsc{leader}_G(t),l,s)$
\label{alg:virtual-leader-decide-leader}
\EndIf
\ElsIf {$w(S,t,(l,1)) > n-d_s$}
\State $v.vote(t) \leftarrow (l,1)$
\Else 
\State $v.vote(t) \leftarrow (l,0)$
\label{alg:virtual-leader-l-c-f-t-z-no-majo}
\EndIf
\ElsIf {$\delta \mod 3 =1$}
\Comment{Coin fixed to $1$}
\If {$w(S,t,(l,1)) > n-d_s$}
\State $v.vote(t) \leftarrow (\oslash,1)$
\label{alg:virtual-leader-decide-non-leader-in round}
\If {$w(S,t,(l,0)) = n$}
\label{alg:virtual-leader-agreement-on-none}
\State $\textsc{long\_chain}(\textsc{leader}_G(t),\oslash,s)$
\label{alg:virtual-leader-decide-non-leader}
\EndIf
\ElsIf {$w(S,t,(l,0)) > n-d_s$}
\State $v.vote(t) \leftarrow (l,0)$
\Else 
\State $v.vote(t) \leftarrow (l,1)$
\EndIf
\ElsIf {$\delta \mod 3 =2$}
\Comment{Genuine coin flip}
\If {$w(S,t,(l,0)) > n-d_s$}
\State $v.vote(t) \leftarrow (l,0)$
\label{alg:virtual-void-leader-genuine-coin-super-majo-0}
\ElsIf {$w(S,t,(l,1)) > n-d_s$}
\State $v.vote(t) \leftarrow (l,1)$
\label{alg:virtual-void-leader-genuine-coin-super-majo-1}
\Else 
\State $b_{coin}\leftarrow lsb(H(x.m))$ for max weight $x\in S$
\State $v.vote(t) \leftarrow (l,b_{coin})$
\label{alg:virtual-void-leader-genuine-coin-no- super-majo}
\EndIf
\EndIf
\EndIf
\EndIf
\EndFor
\end{algorithmic}
\end{algorithm}

Basically, algorithm (\ref{alg:virtual-leader}) computes the votes of a vertex on every local leader election in previous rounds, based on the votes of all members in its safe voting pattern. The special character $\oslash$ is used to indicate, that no actual message could be decided in a round. If the vertex is able to locally decide a leader, the global leader stream is updated, using function (\ref{alg:longest-chain-rule}) as a variation of Nakamoto's longest chain rule. 
 
\begin{algorithm}
\caption{Longest chain rule}
\scriptsize
\label{alg:longest-chain-rule}
\begin{algorithmic}[1]
\Procedure{long\_chain}{set$\langle$uint,\textsc{message}$\rangle$:S,\textsc{message}:m,uint:s}
\If {there is no $(t,l)\in S$ with $t>s$}
\State $S\leftarrow \left( S \backslash \{(t,l)\in S \;|\; t<s\} \right)\cup \{(s,m)\} $
\EndIf
\State \Return $S$ 
\EndProcedure 
\end{algorithmic}
\end{algorithm}

To be more precise, algorithm (\ref{alg:virtual-leader}) computes $v$'s votes in all currently active voting rounds, by looping through the elements of $v.svp$. Each such element indicates a round number and a different stage $\delta$, the latter of which is measured by the position of that round number inside the total order of $v.svp$. 

Any election starts with vertex $v$ proposing its initial vote for a leader in $v$'s own safe voting patter. This is the $\delta=0$ stage of algorithm (\ref{alg:virtual-leader}) and it mimics the initial vote assumption, made in the original $BA^*$ algorithm.

After that, stages $\delta\in\{1,2\}$, basically indicate the two execution steps in Feldman \& Micali's gradecast algorithm $GC$, while all higher stages $\delta\geq 3$ indicate an execution step in Micali's binary agreement protocol $BBA^*$. Of course every such step is entirely virtual and no votes are actually send to other real world processes as explained previously in great detail.

The purpose of the $\delta=1$ stage is to presort all initial votes the vertex received for some round leader message. In fact an actual message $l\neq \oslash$ can become a round leader only, if some vertex receives a super majority of voting weight for that message. If this does not happen, the outcome will be the non-leader $l=\oslash$. Therefore any initial voting weight function has to account for this to ensure liveness.

In stage $\delta=2$ the output of gradecast is transformed into the input of $BBA^*$, to prepare for the local decision either on a single message $l$ or the non-leader message $\oslash$. In this stage an actual leader $l\neq \oslash$ can be proposed only, if a honest majority of voting weight is received for that message.

For any stage with $\delta \geq 3$ and $\delta \text{ mod } 3 = 0$ we are in a 'Coin fixed to zero' round, according to Micali's terms. If a vertex receives voting weight that is in agreement on a vote with zero binary part in such a round, it locally decides a leader and uses the longest chain rule (\ref{alg:longest-chain-rule}) to update the global leader stream. Note however the absence of a stop criteria. This is necessary for the longest chain rule to work properly. We explain this in the next section.

Stage $\delta \geq 3$ and $\delta \text{ mod } 3 = 0$ is analog, but the decision will always be the non leader $\oslash$ message.

For a stage with $\delta \geq 3$ and $\delta \text{ mod } 3 = 2$ we are in a so called 'Genuine coin flip' stage and as usual, no decision is made in such a round. In $BBA^*$ all peers broadcast a unique signature and the least significant bit of the smallest hash of those signature if interpreted as a float, is the same for all participant with probability $2/3$, provided $2/3$ of all peers are honest. 

Our virtualization of such a 'common concrete coin' works as follows: Instead of sending unforgeable signatures, vertices virtually send their own hash and we choose the heaviest of theses hashes to take the least significant bit of it. These hash values are sufficiently unforgeable as the voting weight would drop below the $c_{min}$ threshold if changed, by our tamper-proof assumption. Moreover, we can assume, that the bit $b_{coin}$ is sufficiently random and the same with a non zero probability $p_{coin}$ for all members of the safe voting pattern that contains $v$, because the amount of forking is limited in that voting set. 
\begin{remark}The reader should note, that no termination occurs in any local election. However, once a local leader is decided, every consecutive safe voting pattern, that has such a deciding vertex in its past, will decide the same value due to agreement stability (\ref{thm:agreement-stability}). Hence any actual implementation can stop the local computation for that round and just update the next round accordingly. This is more efficient from an implementation perspective, but the author believes that writing the abstract algorithm without stop criteria is conceptually cleaner.
\end{remark}
\subsection{Total order} 
As time goes by and the Lamport graph grows, more and more round leaders are computed and incorporated into the global leader stream $\textsc{leader}_G(\cdot)$ using procedure (\ref{alg:longest-chain-rule}). We call this function the \textit{longest chain rule}, because it deletes all local leader messages decided previous rounds and keeps those computed in the maximum round number, only. It always chooses the longest chain, so to speak. As we will proof in section (\ref{sec:proof-total-order}), this allows the set of each round $r$ leader to eventually converge to a single element with probability one. Total order is then achieved by topological sorting on the past of appropriate vertices.

The intuition is that the local leader election on a round $r$ never stops, as every new round $s>r$ that has a safe voting patter, recomputes the round $r$ leader. This can be seen as a chain of rounds $s_i$, that all compute the round $r$ leader $(l,s_1)< (l,s_2)< (l,s_3)<\ldots$, but anytime the overall voting weight of such a round exceeds the upper bound $6\cdot d_{s_i}$, additional round leader might appear.

However, every time more then one round $r$ leader appears in a Lamport graph, the chain forks, like $(l,s_1)< \{(l,s_2),(\acute{l},s_2)\}< \{(l,s_3),(\acute{l},s_3)\}< \ldots$. The longest chain rule then selects the maximum hight set of elements in this chain together with all forks that might occur in that set. The reason is, that forks will eventually decay away and a single chain with a single message will asymptotically remain, provided our estimation on the difficulty (\ref{eq:difficulty-bound}) holds. 

Algorithm (\ref{alg:order-loop}) then uses the stream $\textsc{leader}_G(\cdot)$ to compute the total order and as $\textsc{leader}_G(\cdot)$ converges, so does the order. It is executed in an infinite loop and in concurrence to the rest of the system. 
    
\begin{algorithm}
\caption{Order loop}
\label{alg:order-loop}
\scriptsize
\begin{algorithmic}[1]
\Statex \textbf{run the following loop forever} 
\Statex
\Loop { }update order
\State wait for $\textsc{leader}_G(\cdot)$ to change
\State $s\leftarrow$ min round of all changed $\textsc{leader}_G(\acute{s})$
\State $r\leftarrow$ max round of all $\textsc{leader}_G(\acute{r})\neq \emptyset$
\State $v_{l_{s-1}}\leftarrow$ leader in highest round, smaller $s$ in $G$
\For {$s\leq t \leq r$}
\State $n \leftarrow max\{v.total\_position\; | \; v\in Ord_G(v_{l_{t-1}}) \}$ 
\State (randomly) choose $(p,l_t)\in \textsc{leader}_G(t)$
\If {$l_t\neq \oslash$}
\State $\textsc{order}(Ord_G(v_{l_t}),n)$
\Comment $v_{l_t}.m=l_t$
\EndIf
\EndFor
\EndLoop
\Statex
\Statex $Ord_G(v_l)$ past of leader vertex $v_l$ without the past of all leader vertices in previous rounds.
\end{algorithmic}
\end{algorithm}

Every time a round leader appears, or is updated, the algorithm executes a topological sorting algorithm on the past of all future leaders of the smallest updated leader, without reodering the past of previous unchanged leaders. Since that past is invariant between all Lamport graphs by the invariance of the past theorem (\ref{thm:invariant-past}), every process will eventually compute the same total order, provided the leader streams of all honest processes converges. 

Moreover, since we use topological sorting, the generated order will be an extension of the partial causality between massages. 

Efficient topological sorting is known, able to achieve logarithmic run time, if executed concurrently on spacelike vertices. However for the sake of simplicity we use Kahn's algorithm (\ref{alg:kahn-sorting}) as our example, to generate total order in linear runtime. 

\begin{algorithm}
\caption{Total order using Kahn's algorithm}
\label{alg:kahn-sorting}
\scriptsize
\begin{algorithmic}[1]
\Procedure{order}{dag:Ord(v), uint:last}
\State $n\leftarrow last+1$
\State $S\leftarrow$ set of all elements of $Ord(v)$ with no outgoing edges
\While {$S\neq \emptyset$}
\State remove $x$ with highest weight $w(x)$ from $S$
\label{alg:kahn-sorting-line-choose-from-S}
\State $x.total\_position\leftarrow n$
\State $n\leftarrow n+1$
\For {each vertex $y\in Ord(v)$ with edge $e:y\to x$}
\State remove edge $e$ from $Ord(v)$
\If {$y$ has no other outgoing edge}
\State $S\leftarrow S\cup \{y\}$
\EndIf
\EndFor
\EndWhile
\EndProcedure
\Statex
\Statex Kahn's algorithm in its arrow reversed incarnation, since we want to order the past before the future in any Lamport graph.
\end{algorithmic}
\end{algorithm}

Since the weight is invariant among all equivalent vertices and it is practically impossible for two vertices to have the same weight, execution of (\ref{alg:kahn-sorting}) will give the same results in any Lamport graph, which establishes an invariant total order. 

\begin{remark}
Of course sorting by voting weight is just an example. In fact any deterministic function able to decide elements from $S$ in line (\ref{alg:kahn-sorting-line-choose-from-S}) in an invariant way can be used. 
\end{remark}

\subsection{The Crisis protocol}Finally, the overall algorithm works as follows: Member discovery (\ref{alg:discovery-gossip}) and message gossip (\ref{alg:message-gossip}) are executed in infinite loops, concurrently to the rest of the system. Ideally the message sending loop is executed on as many parallel threads as possible. This implies that an overall unbounded amount of new messages arrive over time due to our liveness assumption. In addition each processes may generate messages and write them into its own Lamport graph. 
 
For each new set of messages that pass the integrity check, the Lamport graph is extended by an appropriate set of vertices $V$ that contain those messages. We assume all elements of $V$ to be spacelike and that all vertices in the past of $V$ have already decided round numbers, safe voting patterns and votes. If this is not the case, $V$ can easily be partitioned into sets of spacelike vertices and the protocol is executed on their past first. 

Then, algorithms (\ref{alg:sync-clocks}), (\ref{alg:safe-voting-pattern}) 
and (\ref{alg:virtual-leader}) are executed in that order concurrently on each vertex
from $V$. As these algorithms run, they will update the leader stream $\textsc{leader}_G(\cdot)$ in some way.

In addition, the total order loop (\ref{alg:order-loop}) runs concurrently to the rest of the system and waits for updates of the leader stream. Depending on the actual order algorithm (\ref{alg:kahn-sorting}), additional threads might be required to execute exponentially fast topological ordering algorithms.

\section{Correctness Proof}We show that the Crisis protocol family eventually converts a causal order on messages into a total order on vertices that is asymptotically identical at all nonfaulty processes in the system. In particular we adapt Moser \& Melliar-Smith definition of total order \cite{MM} to our probabilist setting and proof that the following properties hold under the assumptions we make in section (\ref{assumptions}):
\begin{enumerate}
\label{def:byzantine-total-order-axioms}
\item \textit{Probabilistic Termination I}. The probability that a honest process $j$ computes $v.total\_position=i$ for some position $i$ and vertex $v$ increases asymptotically to unity as the number of steps taken by $j$ tends to infinity.
\item \textit{Probabilistic Termination II}. For each message $m$ broadcast by a non byzantine process $j$, the probability that a non byzantine process $k$ places some vertex $v$ with $v.m=m$ in the total order, increases asymptotically to unity as the number of steps taken by $k$ tends to infinity.
\item \textit{Partial Correctness}. The asymptotically convergent total orders determined by any two non byzantine processes are consistent; i.e., if any non byzantine process determines $v.total\_position=i$, then no honest process determines $\acute{v}.total\_position=i$, where $\acute{v}\not\equiv v$.
\item \textit{Consistency}. The total order determined by any non byzantine process is consistent with the partial causality order; i.e. $\acute{v}\leq v$ implies $\acute{v}.total\_position\leq v.total\_position$.
\end{enumerate}

\subsection{Assumptions}
\label{assumptions}
 Our byzantine fault resistant total order is based on the following list of assumptions. 
\begin{enumerate}
\item \textit{Random Oracle Model}. Cryptographic hash functions exist, are collision, first- and second-preimage resistant and behave like random oracles. 
\item \textit{Liveness}. At every moment in time, there are non-faulty processes that participate in the system and every such process must generate further messages that causally follow messages from other nonfaulty process.
\item \textit{Message Dissemination}. If Lamport-graph $G$ of process $j$ contains a vertex $v$ and Lamport-graph $\acute{G}$ of process $k$ does not contain any vertex, equivalent to $v$ and both $j$ and $k$ are honest and participate in the protocol, then there will eventually be a Lamport graph $\tilde{G}$ of process $k$, with $\acute{G}\subset \tilde{G}$ and $v\equiv \tilde{v}$ for some vertex $\tilde{v}\in \tilde{G}$.
\item \textit{Existence of Weight Systems}. A weight system as defined in section (\ref{sec:weight-system}) exists and allows for the definition of a difficulty oracle function $d:\mathbb{N}\to \mathbb{R}^+$, that satisfies (\ref{eq:difficulty-bound}) approximately.
\item \textit{Quorum selector} \& \textit{safe voting pattern}. A quorum selector exists, such that safe voting pattern appear frequently, i.e. the probability $p_r$ that round $r$ has a safe voting pattern is non vanishing.
\item \textit{Initial Vote}. The initial vote function is able to generate vertices with $l\neq \oslash$ in the presorting stage $\delta=2$ of algorithm (\ref{alg:virtual-leader}).
\end{enumerate}

\subsection{Invariance}
Votes are well defined and equal for equivalent vertices among different Lamport graphs. This is the foundation of virtual voting, because any real world process knows, that any other process will compute the same votes with respect to equivalent vertices. In other words, votes are deducible from the causal relation between vertices and we must not send them.

\begin{proposition}[Safe voting pattern invariance] Let $v$ and $\acute{v}$ be two equivalent vertices in Lamport graphs $G$ and $\acute{G}$ respectively. Then $v.svp=\acute{v}.svp$ as well as $v.vote(t)= \acute{v}.vote(t)$ for all $t\in v.svp$.
\end{proposition}
\begin{proof}
Both properties $v.svp$ as well as $v.vote$ depend deterministically on the past of $v$, only. However equivalent vertices have equivalent histories, due to the invariance of the past theorem (\ref{thm:invariant-past}). We therefore proof the statement by strong induction on the number of vertices $|G_v|$ in the histrory of $v$ (which is equal to $|\acute{G}_{\acute{v}}|$). 

For the base case assume that $v$ and $\acute{v}$ are two equivalent vertices in Lamport graphs $G$ and $\acute{G}$ respectively, such that $G_v$ contains $v$ only. Then $\acute{G}_{\acute{v}}$ must contain $\acute{v}$ only and $v.round=0$ as well as $\acute{v}.round=0$ follows. This however implies that no round numbers $s<v.round$ and $\acute{s}< \acute{v}.round$ exist and algorithm (\ref{alg:safe-voting-pattern}) computes the empty total order $v.svp=\emptyset:\emptyset$ as well as $\acute{v}.svp=\emptyset:\emptyset$ in both cases, since the 'if' branch after line (\ref{alg:safe-voting-pattern-first-if}) is not executed. After that, algorithm (\ref{alg:virtual-leader}) executes line (\ref{alg:virtual-leader-return-if-svp-none}) both for $v$ and $\acute{v}$ and we get $v.vote=\bot$ as well as $\acute{v}.vote=\bot$, as no safe voting pattern exist in the past of both $v$ and $\acute{v}$.

For the strong induction step assume that $v$ and $\acute{v}$ are two equivalent vertices in Lamport graphs $G$ and $\acute{G}$ respectively and that $x.svp=\acute{x}.svp$ and $x.vote(t)= \acute{x}.vote(t)$ for all $t\in x.svp$ and equivalent vertices $x$ and $\acute{x}$ in all Lamport graphs $\tilde{G}$ and $\hat{G}$ with $|\tilde{G}_x|< |G_v|$.

If $v.is\_last$ then $\acute{v}.is\_last$ and since the voting sets $S_v(s,k)$ and $S_{\acute{v}}(s,k)$ are equivalent for all $s< v.round=\acute{v}.round$, we know that their overall voting weight must be identical, i.e. $w(S_v(s,k))=w(S_{\acute{v}}(s,k))$. In addition $x.svp=\acute{x}.svp$ as well as $x.vote(u)=\acute{x}.vote(u)$ holds for all $x\in S_v(s,k)$ as well as $\acute{x}\in S_{\acute{v}}(s,k)$ and $u\in x.svp$, by our induction hypothesis, since $|G_x|=|G_{\acute{x}}|<G_v$. This however implies that algorithm (\ref{alg:safe-voting-pattern}) executes the if branch in line (\ref{alg:safe-voting-pattern-first-if}) for $v$, if and only if it executes the same branch for $\acute{v}$. Therefore $v.svp=\acute{v}.svp$.

In case $v.svp=\emptyset$ and $\acute{v}.svp=\emptyset$, algorithm (\ref{alg:virtual-leader}) computes $v.vote=\bot$ as well as $\acute{v}.vote=\bot$ and otherwise executes its for loop on the same round numbers $t$ both for $v$ and $\acute{v}$, using the same $\delta$ in both cases. However, the voting weights of $S_v(s,k)$ and $S_{\acute{v}}(s,k)$ are equal for any vote by our induction hypothesis, since $|G_x|=|\acute{G}_{\acute{x}}|<G_v$. Therefore  algorithm (\ref{alg:virtual-leader}) chooses the same branches both for $v$ and $\acute{v}$, which implies $v.vote(t)=\acute{v}.vote(t)$ for all $t\in v.svp$, since \textsc{initial\_vote} is deterministic and gives the same result on equivalent voting sets and $lsb(H(v.m))=lsb(H(\acute{v}.m))$.

Altogether we get $x.svp=\acute{x}.svp$ and $x.vote=\acute{x}.vote$ for all equivalent $x\in G_v$ and $\acute{x}\in\acute{G}_{\acute{v}}$, which proofs the proposition by strong induction.
\end{proof}

\subsection{Virtual Leader Election}
We proof that the virtual leader election algorithm (\ref{alg:virtual-leader}) eventually decides a set of round leader with probability one. As algorithm (\ref{alg:virtual-leader}) is an adaptation of Chen, Feldman \& Micali's algorithm $BA^*$, we follow their ideas and divide the proof into two subproofs, the first of which shows the graded consensus properties and the second of which proof the binary consensus part.

\begin{remark}
In what follows we will frequently say that some vertex $v$ executes the virtual leader election algorithm (\ref{alg:virtual-leader}). This in agreement with our line of thought, because a vertex $v$ represents a virtual process $v.id$. But of course algorithm (\ref{alg:virtual-leader}) is executed by some real world process with input $(v,G)$, where $G$ is a Lamport graph that contains $v$.
\end{remark}

Consensus protocols are based on a property called agreement, which basically means that all honest processes hold the same value. However when it comes to weighted consensus it might not be appropriate to distinguish between a honest and a faulty process, but to talk about honest and faulty weight instead. We therefore say that the honest voting weight is in agreement on some vote, if all but a possibly byzantine amount of weight agrees on that vote. 
\begin{definition}[Agreement]Let $v$ be a vertex with $|v.svp|\geq 3$, $\{r>s\}\in v.svp$ the largest two elements and $S_v(r,k)$ the safe voting pattern of $v$. We then say that the members of $S_v(r,k)$ are in \textbf{agreement} on some round $t\in v.svp\backslash\{r,s\}$ vote $(l,b)$, if the voting set $S_x(s,k)$ of each such member $x\in S_v(r,k)$ has a super majority of voting weight for $(l,b)$, that is the inequality $w(S_x(s,k),t,(l,b))> w(S_x(s,k))\ominus d_s$ holds.
\end{definition}
As the following corollary shows, our definition of agreement, immediately implies, that all members of a voting set compute the same vote for any agreed on value.

\begin{corollary}
\label{thm:agreement-implies-all}
Let $v$ be a vertex with $|v.svp|\geq 3$ and $\{r>s\}\subset v.svp$ the largest two elements, such that the members of $S_v(r,k)$ are in agreement on some round $t\in v.svp\backslash\{r,s\}$ vote $(l,b)$. Then $x.vote(t)=y.vote(t)$ for all $x,y\in S_v(r,k)$.
\end{corollary}
\begin{proof}
Each member $x\in S_v(r,k)$ executes algorithm (\ref{alg:virtual-leader}) to compute its own votes. Since $t<s<r$, we know $t<r-1$, which implies that the 'Initial leader proposal' branch $\delta=0$ is never executed for any $x\in S_v(r,k)$. But since a super majority of voting weight from $S_x(s,k)$ votes for $(l,b)$, line (\ref{alg:virtual-leader-l-for-positive-delta}) computes the same $l$ again and one of the following branches decides the same $\acute{b}\in\{\bot,0,1\}$ for all $x\in S_v(v,k)$. Hence $x.vote(t)=y.vote(t)$ for all $x,y\in S_v(r,k)$.
\end{proof}
\subsubsection{Graded Consensus}
We start the correctness proof with a series of propositions, that basically show that the first three branches (i.e. $\delta\in\{0,1,2\}$) of our virtual leader election (\ref{alg:virtual-leader}) are nothing but an adaptation of Feldman \& Micali's graded consensus, but executed in three consequitive safe voting patterns. Our proofs are strongly influenced by the approach taken in \cite{FM}. 

\begin{proposition}[Initial super majority]
\label{thm:initial-super-majority}
Let $v$ be a vertex with $|v.svp|\geq 3$ and $\{r>s>t\}\subset v.svp$ the three maximum elements from $v.svp$. If there is a member $x\in S_v(r,k)$ that receives a super majority of voting weight $w(S_x(s,k),t,(l,\bot))$ from its voting set $S_x(s,k)$ for some round $t$ initial vote $(l,\bot)$, there can not be a member $y\in S_v(r,k)$ that receives a super majority of voting weight from its voting set $S_y(s,k)$ for some round $t$ vote $(\acute{l},\bot)$ with $l\neq \acute{l}$.
\end{proposition}
\begin{proof}The statement follows from the properties of a safe voting pattern. To see this in detail, first observe that if algorithm (\ref{alg:virtual-leader}) is executed from $x\in S_v(r,k)$, the $\delta=1$ branch is used to compute $x.vote(t)$, since $x.svp=v.svp\backslash \{r\}$ by proposition (\ref{thm:svp-of-svp-member}) and therefore $d_{x.svp}(s,t)=1$. This however implies that the binary part of  $x.vote(t)$ is undecided for all $x\in S_v(r,k)$.

We proof the statement by contradiction. Suppose that there is another member $y\in S_v(r,k)$ that receives a super majority of voting weight $w > w(S_y(s,k))\ominus d_{s}$ for some vote $(\acute{l},\bot)$ with $l\neq \acute{l}$ from its voting set $S_y(s,k)$. Then $x$ must receive more then $w(S_y(s,k))\ominus 2\cdot d_s$  voting weight for $(\acute{l},\bot)$, since $S_v(r,k)$ is a safe voting pattern, which implies $|w(S_y(s,k),t,(\acute{l},\bot))\ominus w(S_x(s,k),t,(\acute{l},\bot))|\leq d_s$. 

Moreover since the overall voting weight of $S_y(s,k)$ is strictly larger then $3\cdot d_s$, $x$ must receive more then $d_s$ voting weight for $(\acute{l},\bot)$ and at the same time more then $w(S_x(s,k))\ominus d_s$ voting weight for $(l,\bot)$, which is a contradiction.

To see that, let $S_x^{(l,\bot)}$ be the set of all members from $S_x(s,k)$ that vote for $(l,\bot)$ and $S_x^{(\acute{l},\bot)}$ the set of members from $S_x(s,k)$ that vote for $(\acute{l},\bot)$. Then
\begin{align*}
w(S_x^{(\acute{l},\bot)} \cap S_x^{(l,\bot)}) = \\
w(S_x^{(\acute{l},\bot)})\oplus w(S_x^{(l,\bot)})\ominus
w(S_x^{(\acute{l},\bot)} \cup S_x^{(l,\bot)}) > \\
d_s \oplus w(S_x(s,k))\ominus d_s \ominus w(S_x^{(\acute{l},\bot)} \cup S_x^{(l,\bot)})=\\
w(S_x(s,k)) \ominus w(S_x^{(\acute{l},\bot)} \cup S_x^{(l,\bot)})\geq 0
\end{align*}
This means that there are members of $S_x(s,k)$ that have a vote both for $(l,\bot)$ and $(\acute{l},\bot)$ in round $t$, which is a contradiction, since each vertex has a single vote in any round only. 
\end{proof}
\begin{proposition}[Message presorting]
\label{thm:message-presorting}
Let $v$ be a vertex with $|v.svp|\geq 3$ and $\{r>s>t\}\subset v.svp$ the three maximum elements from $v.svp$. Then there is a message $l$ and each member of $S_v(r,k)$ has a round $t$ vote either for $(l,\bot)$ or $(\oslash,\bot)$ and no other round $t$ votes appear in $S_v(r,k)$.
\end{proposition}
\begin{proof}
Let $x\in S_v(r,k)$. Then algorithm (\ref{alg:virtual-leader}) computes $\delta=1$ if executed by $x$ and $x$ either received a super majority of initial voting weight for some round $t$ vote $(l,\bot)$ from its voting set $S_x(s,k)$ or it does not. In the first case $x.vote(t)=(l,\bot)$ and in the second $x.vote(t)=(\oslash,\bot)$. Now suppose that $x$ and $y$ are both members of $S_v(r,k)$, that both received a super majority of voting weight for some round $t$ vote $(l,\bot)$ and $(\acute{l},\bot)$, respectively. Then the previous proposition (\ref{thm:initial-super-majority}) implies $l=\acute{l}$.
\end{proof}

\begin{proposition}[Graded Agreement]
\label{thm:graded-agreement}
Let $v$ be a vertex with $|v.svp|\geq 4$, $\{r>s>t>u\}\subset v.svp$ the four maximum elements from $v.svp$ and let $x$ and $y$ be two members of $S_v(r,k)$. If $x$ has round $u$ vote  $(l,1)$ or $(l,0)$ for some message $l\neq \oslash$ and $y$ has round $u$ vote $(\acute{l},1)$ or $(\acute{l},0)$ for some message $\acute{l}\neq \oslash$, then $l=\acute{l}$.
\end{proposition}
\begin{proof}This is our adaptation of the $g_i,g_j>0 \Rightarrow v_i=v_j$ property in the definition of graded consensus and a consequence of the previous proposition.

To see that, first observe that if algorithm (\ref{alg:virtual-leader}) is executed from $x\in S_v(r,k)$, the $\delta=2$ branch is used to compute $x.vote(u)$, since $x.svp=v.svp\backslash \{r\}$ by proposition (\ref{thm:svp-of-svp-member}) and therefore $d_{x.svp}(s,u)=2$. 

Now, if $x$ votes for $(l,0)$ or $(l,1)$ in round $u$ with $l\neq \oslash$, it must have received more then $d_s$ voting weight for $(l,\bot)$ from its voting set $S_x(s,k)$ and since $S_v(r,k)$ is a safe voting patter, we know $|w(S_x(s,k),(l,\bot))-w(S_y(s,k),(l,\bot))|\leq d_s$ for all members $y\in S_v(r,k)$. Hence each member of $S_v(r,k)$ must have received at least some voting weight for $(l,\bot)$. 

This implies that there must be a member of $y$'s safe voting pattern $S_y(s,k)$ that has a round $u$ vote $(l,\bot)$. But from the previous proposition (\ref{thm:message-presorting}) we know, that then no other member in $y$'s safe voting pattern $S_y$ can vote for some actual message $\acute{l}\neq l$. Therefore if $y$ does not vote $(\oslash,\bot)$, it must have voted $(l,0)$ or $(l,1)$.
\end{proof}

\begin{proposition}[Bounded grading]
\label{thm:bounded-grading} 
Let $v$ be a vertex with $|v.svp|\geq 4$, $\{r>s>t>u\}\subset v.svp$ the four maximum elements from $v.svp$ and let $x$ be a members of $S_v(r,k)$ with $x.vote(u)=(l,0)$ for some message $l\neq\oslash$. Then there can not be a member $y\in S_v(r,k)$ with $y.vote(u)=(\oslash,1)$ 
\end{proposition}
\begin{proof}
This is our adaptation of the $|g_i-g_j|\leq 1$ property in the definition of graded consensus. To start, observe that if algorithm (\ref{alg:virtual-leader}) is executed from $x\in S_v(r,k)$, the $\delta=2$ branch is used to compute $x.vote(u)$, since $x.svp=v.svp\backslash \{r\}$ by proposition (\ref{thm:svp-of-svp-member}) and therefore $d_{x.svp}(s,u)=2$.

Now, for $x$ to compute $x.vote(u)=(l,0)$, $l$ must be an actual message, i.e $l\neq\oslash$ and $x$ must have received a super majority of voting weight for $(l,\bot)$ from its voting set $S_x(s,k)$.

But then every other member $y\in S_v(r,k)$ must receive strictly more then $d_{s}$ voting weight for $(l,\bot)$, since $S_v(r,k)$ is a safe voting pattern, which implies $w(S_y(s,k))> 3\cdot d_{s}$ as well as 
$|w(S_x(s,k),(l,\bot))\ominus w(S_y(s,k),(l,\bot))|\leq d_{s}$. This however implies, that $y$'s execution of (\ref{alg:virtual-leader}) can not compute $y.vote(u)=(\oslash,1)$.
\end{proof}

\begin{proposition}[Graded consistency]
Let $v$ be a vertex with $|v.svp|\geq 4$, $\{r>s>t>u\}\subset v.svp$ the largest four elements and let there be a vertex $\acute{v}\in S_v(r,k)$, such the members 
$x\in S_{\acute{v}}(s,k)$ are in agreement on a round $u$ vote $(l,\bot)$. Then each member $y\in S_v(r,k)$ computes its round $u$ vote as $y.vote(u)=(l,0)$.
\end{proposition}
\begin{proof}
For a first orientation, observe that the execution of algorithm (\ref{alg:virtual-leader}) from $\acute{v}\in S_v(r,k)$, uses the $\delta=2$ branch to compute $\acute{v}.vote(u)$, since $\acute{v}.svp=v.svp\backslash \{r\}$ by proposition (\ref{thm:svp-of-svp-member}) and therefore $d_{\acute{v}.svp}(s,u)=2$. The same reasoning shows that the $\delta=1$ branch is used to compute $x.vote(u)$ if algorithm (\ref{alg:virtual-leader}) is executed from any $x\in S_{\acute{v}}(s,k)$. 

Since the members of $S_{\acute{v}}(s,k)$ are in agreement on a round $u$ vote $(l,\bot)$, by definition each member $x\in S_{\acute{v}}(s,k)$ receives a super majority of initial voting weight for $(l,\bot)$ from its voting set $S_x(t,k)$, which implies $x.vote=(l,\bot)$ for all $x\in S_v(s,k)$ by corollary (\ref{thm:agreement-implies-all}).

Then $\acute{v}$ receives all voting weight $w(S_{\acute{v}}(s,k))$ for $(l,\bot)$ and no voting weight $w(S_{\acute{v}}(s,k),(\acute{l},\bot))=0$ for any other vote $(\acute{l},\bot)$ with $l\neq \acute{l}$. However since $S_v(r,k)$ is a safe voting pattern, each member $y\in S_{v}(r,k)$ receives at most $d_r$ voting weight for any $(\acute{l},\bot)$ other then $(l,\bot)$ from its voting set $S_y(s,k)$, which implies, that the overall voting weight $y$ receives for $(l,\bot)$ must be at least $w(S_y(s,k))\ominus d_s$, since each member of $S_y(s,k)$ has exactly one round $u$ vote. But then $y$'s execution of (\ref{alg:virtual-leader}) gives $\delta=2$ and then $y.vote(u)=(l,0)$. 
\end{proof}

\subsubsection{Binary byzantine agreement} As we have seen in the previous section, the $\delta\in\{0,1,2\}$ branches of algorithm (\ref{alg:virtual-leader}) are an adaptation of Feldman \& Micali's gradecast algorithm. In this section, we proof that the $\delta\geq 3$ branches simulate Micali's binary byzantine agreement protocol $BBA^*$, if we, for a moment, consider the binary part $(\cdot,b)$ of any vote $(l,b)$ only. Our proofs are strongly influenced by Micali's original ideas as provided in \cite{SM}.  

\begin{proposition}[Binary quorum intersection]
\label{thm:quorum-intersection} Let $v$ be a vertex with $|v.svp|\geq 5$, 
$\{r>s\}\subset v.svp$ the largest two elements of $v.svp$ and $x\in S_v(r,k)$ a member that receives a super majority of voting weight $w> w(S_{x}(s,k))-d_{s}$ for some round $u$ vote $(\cdot,0)$ with $d_{x.svp}(s,u)\geq 4$ from its voting set $S_{x}(s,k)$. Then there is no member $y\in S_v(r,k)$, that receives a super majority of voting weight $ w(S_{y}(s,k))-d_{s}$ for a round $u$ vote $(\cdot,1)$ from its voting set $S_{y}(s,k)$ and vice versa. 
\end{proposition}
\begin{proof}
First observe that the execution of algorithm (\ref{alg:virtual-leader}) from any $y\in S_x(s,k)$, uses a $\delta\geq 3$ branch to compute $y.vote(u)$, since $y.svp=x.svp\backslash \{s\}$ by proposition (\ref{thm:svp-of-svp-member}) and therefore $d_{y.svp}(t,u)\geq 3$ for $t=max(y.svp)$. Therefore the binary part of $y.vote(u)$ is decided for any $y\in S_x(s,k)$.

We proof the theorem by contradiction and assume that there is a member $x$ of $S_v(r,k)$, that receives a super majority of voting weight $ w > w(S_{x}(s,k))\ominus d_{s}$ for a round $u$ vote $(\cdot,0)$ from its voting set $S_{x}(s,k)$ and another member $y\in S_v(r,k)$ that received a super majority of voting weight $w(S_{y}(s,k))\ominus d_{s}$ for a vote $(\cdot,1)$ in the same round through its voting set $S_{y}(s,k)$.

Since $S_v(r,k)$ is a safe voting pattern, both voting sets $S_{x}(s,k)$ and $S_{y}(s,k)$ have an overall weight strictly larger then $3\cdot d_s$. Moreover $|w(S_{x}(s,k), (\cdot,0))\ominus w(S_{y}(s,k),(\cdot,0))|< d_s$ implies that $y$ must have received strictly more then $w(S_{x}(s,k))\ominus 2\cdot d_s$ voting weight for $(\cdot,0)$. 
 
Now, let $S_y^{0}$ and $S_y^{1}$ be the subsets of $S_y(s,k)$ through which $y$ received votes for $(\cdot,0)$ and $(\cdot,1)$, respectively. Then $S_y^{*}:= S_y^{0} \cup S_y^{1}$ is again a subset of $S_y(s,k)$. If we use the identity $w(S)=w(S_{1})\oplus w(S_{2}) \circleddash w(S_{1}\cap S_{2})$ for the weights of a cover $S_1$ and $S_2$ of a set $S$ we get
\begin{align*}
w(S_y^{1}\cap S_y^{0})=&\\ 
w(S_y^{1}) \oplus w(S_y^{0}) \circleddash w(S^{*}_{y}) > &\\
w(S_y(s,k))\ominus d_s \oplus w(S_x(s,k))\ominus 2d_s \circleddash w(S^{*}_{y})=&\\ 
\left(w(S_y(s,k))\ominus w(S_{y}^{*})\right) \oplus \left(w(S_x(s,k))\ominus 3d_s \right)> &\; 0
\end{align*}
since $S_y^{*}$ is a subset of $S_y(s,k)$ and the weight of $S_x(s,k)$ is strictly larger then $3d_s$. But no vertex can vote both for $(\cdot,0)$ and $(\cdot,1)$ in the same round. Hence we arrive at a contradiction. The proof for the vice versa case is exactly analog.
\end{proof}

\begin{proposition}
\label{thm:will-decide-eventually}
Let $v$ be a vertex with $|v.svp|\geq 5$, $\{r>s\}\subset v.svp$ the largest two elements from $v.svp$ and $p_{coin}$ the probability that $b:=lsb(H(\acute{v}_x.m))$ is the same for all $x\in S_v(r,k)$ and maximum weight vertex $\acute{v}_x\in S_x(s,k)$. If there is an element $u\in v.svp$ with $d_{v.svp}(s,u)\geq 3$ and  $d_{v.svp}(s,u)\text{ mod } 3=2$, then, with probability at least $p_{coin}/2$, all members $x\in S_v(r,k)$ will have the same vote $x.vote(u)=(\cdot,b)$ for a binary value $b\in\{0,1\}$.
\end{proposition}
\begin{proof}
Since $x.svp=v.svp\backslash \{r\}$ for every $x\in S_v(r,k)$ by proposition (\ref{thm:svp-of-svp-member}), the requirement $d_{v.svp}(s,u) \text{ mod } 3 =2$ implies that the computation of $x$'s round $u$ vote in algorithm (\ref{alg:virtual-leader}) executes the $\delta \text{ mod } 3 =2$  branch, e.g. the genuine-coin-flip stage. The quorum intersection theorem (\ref{thm:quorum-intersection}) then induces the following five exclusive cases:

1.) Every member $x\in S_v(r,k)$ receives a super majority of votes $(l,0)$ for some message $l$ and computes $x.vote(t)=(l,0)$ in line (\ref{alg:virtual-void-leader-genuine-coin-super-majo-0}). Hence every member votes $x.vote(\cdot,0)$

2.) Every member $x\in S_v(r,k)$ receives a super majority of votes $(l,1)$ for some message $l$ and computes $x.vote(s)=(l,1)$ in line (\ref{alg:virtual-void-leader-genuine-coin-super-majo-1}). Hence every member votes $x.vote(\cdot,1)$

3.) No member $x\in S_v(r,k)$ receives a super majority, neither for $(l,0)$ nor for $(\acute{l},1)$. Hence all members of $S_v(r,k)$ compute their vote as $(\cdot, b)$ in line (\ref{alg:virtual-void-leader-genuine-coin-no- super-majo})
where $b$ is the least significant bit $b:=lsb(H(\acute{v}_x.m))$ of the vertex $\acute{v}_x\in S_x(s,k)$ that has the highest voting weight in $S_x(s,k)$. Then with probability $p_{coin}$, agreement will hold on $(\cdot,b)$, as the probability of $b$ being the same for all $x\in S_v(r,k)$ is assumed to be $p_{coin}$.

4.) Some members $x\in S_v(r,k)$ receive a super majority of voting weight for some $(l,0)$ and some neither receive a super majority for $(\acute{l},0)$ nor for $(\tilde{l},1)$. Let $S_0\subset S_v(r,k)$ be the set of members that receives a super majority of voting weight for $(l,0)$ and $S_{b}$ the set of members that receives no super majority at all. Then all processes in $S_{0}$ execute line (\ref{alg:virtual-void-leader-genuine-coin-super-majo-0}) and vote $(\cdot,0)$ and all processes in $S_{b}$ execute line (\ref{alg:virtual-void-leader-genuine-coin-no- super-majo}) and vote $(\cdot,b)$. Hence with probability at least $p_{coin}/2$, every member $x\in S_v(s,k)$ votes $x.vote(t)=(\cdot, 0)$.

5.) Some members $x\in S_v(r,k)$ receive a super majority of voting weight for some $(l,1)$ and some  neither receive a super majority for $(\acute{l},0)$ nor for $(\tilde{l},1)$. The argumentation is then analog to the previous situation.

From the quorum intersection theorem (\ref{thm:quorum-intersection}), we know that it is impossible for two members of the same safe voting pattern to receive super majorities both for $(\cdot,0)$ and $(\cdot,1)$. Hence the previous five case are exclusive and the proposition follows.  
\end{proof}

\begin{proposition}[Agreement stability]
\label{thm:agreement-stability}
Let $v$ be a vertex with $|v.svp|\geq 5$, $r=max(v.svp)$ the maximum element from $v.svp$ and suppose that there is an element $u\in v.svp$ with $d_{v.svp}(r,u)\geq 3$ such that the members $x\in S_v(r,k)$ are in agreement on some round $u$ binary vote $(\cdot,b)$ with $b\in\{0,1\}$. If $\acute{v}$ is another vertex with $\acute{v}.svp\neq \emptyset$, $q=max(\acute{v}.svp)$ and $v\in S_{\acute{v}}(q,k)$, then every member $y\in S_{\acute{v}}$ has the same round $u$ vote $y.vote(u)=(\cdot,b)$, too.
\end{proposition}
\begin{proof}We proof the proposition for $(\cdot,0)$. The situation for $(\cdot,1)$ is analog. Since the members of $S_v(r,k)$ are in agreement on a round $u$ binary vote $(\cdot,0)$, corollary (\ref{thm:agreement-implies-all}) implies, that every member $x\in S_v(r,k)$ computes $x.vote(u)=(\cdot,b)$. Hence all voting weight of the safe voting pattern votes for $(\cdot,0)$, e.g. $w(S_v(r,k),u,(\cdot,0))=w(S_v(r,k))$ as well as $w(S_v(r,k),u, (\cdot,1))=0$ holds, as no vertex can have more then one vote in a round. But since $S_{\acute{v}}(q,k)$ is a safe voting pattern, we know $|w(S_v(r,k),u, (\cdot,1))\ominus w(S_y(r,k),u, (\cdot,1))|< d_{r}$ for all $y\in S_{\acute{v}}$, which implies $w(S_y(r,k),u, (\cdot,1))< d_r$. However each member of $S_y(r,k)$ either votes for $(\cdot,0)$ or $(\cdot,1)$, since ever member has exactly one vote in a round. This implies $w(S_y(r,k),u, (\cdot,0))= w(S_y(r,k))\ominus w(S_y(r,k),(\cdot,1))< w(S_y(r,k))\ominus d_r$. Hence $y$ receives a super majority of voting weight for $(\cdot,0)$ and therefore votes $(\cdot,0)$. This is true for all $y\in S_{\acute{v}}(q,k)$.
\end{proof}

\begin{proposition}
Let $v$ be a vertex with $|v.svp|\geq 5$, $\{r>s\}\subset v.svp$ the highest two elements and let $u\in v.svp$ be a round number with $d_{v.svp}(r,u)> 3$, such that there is a member $x\in S_v(r,k)$ that receives a super majority of voting weight $w > w(S_x(s,k))\ominus d_s$ from its voting set, for a round $u$ vote $(\cdot,b)$ with $b\in\{0,1\}$ and $d_{v.svp}(s,u) \mod 3 = b$. Then all members $y\in S_{v}(r,k)$ vote $y.vote(u)=(\cdot,b)$.
\end{proposition}
\begin{proof}
We proof the proposition for $(\cdot,0)$. The situation for $(\cdot,1)$ is analog.

In that case every member $x\in S_v(r,k)$ executes the $\delta \text{ mod } 3 =0$, i.e. the coin-fixed-to-zero branch of algorithm (\ref{alg:virtual-leader}). Since $x$ received a super majority of voting weight $w > w(S_x(s,k))\ominus d_s$ from its voting set $S_x(s,k)$ for a round $u$ vote $(\cdot,0)$ and $S_v(r,k)$ is a safe voting pattern, no other member $y\in S_v(r,k)$ can receive a super majority of voting weight for a vote $(\cdot,1)$ due to the binary quorum interesection theorem (\ref{thm:quorum-intersection}). However this implies, that each member of $y\in S_v(r,k)$ computes its vote as $(\cdot,0)$ either according to line 
\ref{alg:virtual-leader-l-c-f-t-z-super-majo} or line 
\ref{alg:virtual-leader-l-c-f-t-z-no-majo}. 
\end{proof}

\begin{proposition}[Eventual Agreement]
\label{thm:eventuel-binary-agreement}
Suppose that the probability for the appearance of new rounds and safe voting pattern is not zero. Let $s$ be a round number, such that there is a safe voting pattern $S_{\acute{v}}(s,k)$ for some vertex $\acute{v}$ in round $s$. Then, with probability one, there will be a vertex $v$ with $r=max(v.svp)$ and $s\in v.svp$, such that all members of $S_v(r,k)$ will be in agreement on the binary part of their round $s$ votes, i.e. $x.vote(s)=(\cdot,b)$ 
holds for all $x\in S_v(r,k)$. 
\end{proposition}
\begin{proof}
As the probability of new rounds to appear is not zero and safe voting patterns will appear at least in some of these rounds, there will be vertices $v$, with $v.svp\neq \emptyset$, $s\in v.svp$ and $d_{v.svp}(max(v.svp),s) \mod 3 = 2$. But then proposition (\ref{thm:will-decide-eventually}) implies that the probability to reach agreement on some round $s$ vote in $S_v(r,k)$ is not zero. Since there is an unbounded amount of those vertices, agreement holds eventually with probability one. 
\end{proof}

\subsubsection{Virtual leader agreement}
\begin{proposition}[Eventual Agreement]
\label{thm:eventuell-agreement}
Suppose that the probability for the appearance of new rounds and safe voting pattern is not zero. Let $s$ be a round number, such that there is a safe voting pattern $S_{\acute{v}}(s,k)$ for some vertex $\acute{v}$ in round $s$. Then, with probability one, there will be a vertex $v$ with $r=max(v.svp)$ and $s\in v.svp$, such that all members of $S_v(r,k)$ will be in agreement on a round $s$ vote, i.e. $x.vote(s)=(l,b)$ holds for all $x\in S_v(r,k)$ and message $l$. 
\end{proposition}
\begin{proof}
Due to proposition (\ref{thm:eventuel-binary-agreement}), we know that with probability one, there will be a vertex $v$ with $r=max(v.svp)$ and $s\in v.svp$, such that all members of $S_v(r,k)$ will be in agreement on the binary part of a round $s$ vote, i.e. $x.vote(s)=(\cdot,b)$ holds for all $x\in S_v(r,k)$ and some $b\in\{0,1\}$. 

If binary agreement holds on $(\cdot, 1)$, that is $x.vote(s)=(\cdot,1)$ for all $x\in S_v(r,k)$, then line (\ref{alg:virtual-leader-decide-non-leader-in round}) will be executed by every member of $S_v(r,k)$, hence each such member computes $x.vote(s)=(\oslash,b)$ and agreement holds on $\oslash$.

If binary agreement holds on $(\cdot,0)$, then $|v.svp|\geq 5$ and there is a $t\in v.svp$ such that $d_{v.svp}(t,s)=3$. Then there is a vertex $\tilde{v}$ and a safe voting patter $S_{\tilde{v}}(t,k)$ in the past of $v$, such that at least one member must have received a super majority of voting weight for some vote $(l,\bot)$ with $l\neq \oslash$, because otherwise, all members of $S_{\tilde{v}}(t,k)$ would be in agreement on $(\cdot,1)$ and by proposition (\ref{thm:agreement-stability}) stay in agreement on that vote, which contradicts our assumption, that agreement holds on $(\cdot,0)$.

Hence proposition (\ref{thm:bounded-grading}) implies, that no member of $S_{\tilde{v}}(t,k)$ votes $(\oslash,1)$ and proposition (\ref{thm:graded-agreement}) then implies that all members of of $S_{\tilde{v}}(t,k)$ either vote $(l,0)$ or $(l,1)$ for the same message $l$. In any case all members of that round are in agreement on the message $l$. Therefore $l$ always receives the most voting weight in consecutive rounds (simply because there is no other choice) and hence agreement continous to hold on $l$, which implies that all members of 
$S_v(r,k)$ compute $x.vote(s)=(l,0)$
\end{proof}
\begin{proposition}[Agreement stability]
\label{thm:general-agreement-stability}
Let $v$ be a vertex with $v.svp\neq\emptyset$, $r=max(v.svp)$ the largest element from $v.svp$ and let there be an element $t\in v.svp$ with $d_{v.svp}(r,t)\geq 3$ and the members $x\in S_v(r,k)$ are in agreement on some message $l$, i.e $x.vote(t)=(l,b)$ with $b\in\{0,1\}$. If $\acute{v}$ is another vertex with $\acute{v}.svp\neq \emptyset$, $q=max(\acute{v}.svp)$ and $v\in S_{\acute{v}}(q,k)$, then every member $y\in S_{\acute{v}}$ has a vote $y.vote(t)=(l,b)$ in round $t$, too.
\end{proposition}
\begin{proof}This follows from the binary agreement stability (\ref{thm:agreement-stability}) and proposition (\ref{thm:eventuell-agreement}).
\end{proof}

\subsection{Total Order}
\label{sec:proof-total-order}
\subsubsection{Leader stream convergence}
\begin{proposition}
\label{thm:leader-contains-elements}
Suppose that the probability for the appearance of new rounds and safe voting pattern is not zero, let $j\in \Pi$ be a honest process and $r$ a round number. Then $j$ will eventually have a Lamport graph $G$, such that the set $\textsc{leader}_G(r)$ is not empty.
\end{proposition}
\begin{proof} If there will never be a safe voting pattern in round $r$, algorithm (\ref{alg:virtual-leader}) will eventually execute line (\ref{alg:virtual-void-leader-if-svp-none}) for some vertex and insert $(r,\oslash)$ into $\textsc{leader}_G(r)$. If on the other hand $r$ has a safe voting patter, proposition (\ref{thm:eventuell-agreement}) implies that with probability one there will eventually be a vertex that has a safe voting pattern, such that all members of that pattern are in agreement on a round $r$ leader. In that case, execution of (\ref{alg:virtual-leader}) will enter line (\ref{alg:virtual-leader-decide-leader}) or line (\ref{alg:virtual-leader-decide-non-leader}) and therefore elements are inserted into $\textsc{leader}_G(r)$.
\end{proof}

\begin{proposition}
\label{thm:each-local-leader-adds-3ds}
Let $j\in \Pi$ be a honest process that has a Lamport graph $G$, such that there are $n$ different elements in the set $\textsc{leader}_G(r)$. Then there is a round $s$ in $G$ with an overall amount of voting weight $w_s^G$  strictly larger then $3\cdot n\cdot d_s$.
\end{proposition}
\begin{proof}
First of all, function (\ref{alg:longest-chain-rule}) ensures, that all elements in $\textsc{leader}_G(r)$ always have the same round number, because a new element $(t,l)$ is inserted only, if there are no elements $(\acute{t},\acute{l})\in\textsc{leader}_G(r)$, that have higher deciding rounds $\acute{t}>t$. Moreover, once an element is inserted, all elements with lower deciding rounds are deleted. This implies that the massage part of different elements from $\textsc{leader}_G(r)$ must differ, but the round parts are always the same.

Thats being said, we proof the proposition in case there are two different elements in $\textsc{leader}_G(r)$ only. The general argumentation is analog. To see that, let  $(l,s)$ and $(\acute{l},s)$ be different elements of $\textsc{leader}_G(r)$. Then we know that there must be two different vertices $v$ and $\acute{v}$, that both have round $s$ safe voting patterns $S_v(s,k)$ and  $S_{\acute{v}}(s,k)$, such that $v$'s execution of algorithm (\ref{alg:virtual-leader}) inserted $(l,s)$ and $\acute{v}$'s execution inserted $(\acute{l},s)$ into $\textsc{leader}_G(r)$. 

However due to the execution of line (\ref{alg:virtual-leader-agreement-on-l}), or (\ref{alg:virtual-leader-agreement-on-none}), $v$'s safe voting pattern $S_v(s,k)$ is in agreement on $(l,b)$ and $\acute{v}$'s safe voting pattern $S_{\acute{v}}(s,k)$ is in agreement on $(\acute{l},b)$, e.g. all members $x\in S_v(s,k)$ vote $x.vote(r)=(l,b)$ and all members $\acute{x}\in S_{\acute{v}}(s,k)$ vote $\acute{v}.vote(r)=(\acute{v},b)$ for some $b\in\{0,1\}$. But since any vertex has a single vote in any round only, both voting sets must be disjoint. However $S_v(s,k)$ as well as $S_{\acute{v}}(s,k)$ are safe voting patterns and each has an overall amount of voting weight strictly larger then $3\cdot d_{s}$.
\end{proof}

\begin{theorem}[Leader convergence]
\label{thm:leader-stream-convergence}
Suppose that the probability for the appearance of new rounds and safe voting pattern is not zero and let $j\in \Pi$ be a honest process. Then $j$ will have a series of Lamport graphs $G(t)$, such that the series of sets $\textsc{leader}_{G(t)}(r)$ converges to contain a single element only. 
\end{theorem}
\begin{proof}Since new rounds appear, $j$ will obtain a stream of messages, that extend the current Lamport graph. The time indexed Lamport graphs can therefore be seen as a sequence, such that each consecutive graph contains strictly more elements then the previous one. Despite the fact, that time is a continuous index. The theorem then follows from proposition (\ref{thm:each-local-leader-adds-3ds}), our assumption (\ref{eq:difficulty-bound}) on the boundary of the difficulty oracle and agreement stability.

To see this in detail, we proof the theorem by contradiction and assume that $\textsc{leader}_{G(t)}(r)$ does not converge to a single element for $t\to\infty$. Then proposition (\ref{thm:leader-contains-elements}) implies that there is a parameter $t_0$, such that each set $\textsc{leader}_{G(t)}(r)$ contains at least two elements for all $t>t_0$.

Let $t_1$ be a time parameter, such that there are at least two elements $(s,l)$ and $(s,\acute{l})$ in $\textsc{leader}_{G(t_1)}(r)$. Proposition (\ref{thm:each-local-leader-adds-3ds}) then implies that each element is decided by execution of algorithm (\ref{alg:virtual-leader}) from a vertex $v_1$ and a vertex $\acute{v}_1$ both of which have disjoint safe voting pattern and the overall voting weight of round $s$ is strictly larger in the Lamport graph $G(t_1)$. 

Now since new rounds and safe voting patterns appear forever there must be a time $t_2$ and
two vertices $v_2$ and $\acute{v}_2$ that have safe voting pattern $S_{v_2}(s_1,k)$ and $S_{\acute{v}_2}(\acute{s}_1,k)$, such that $v_1$ is in the safe voting pattern of $v_2$ and $\acute{v}_1$ is in the safe voting pattern of $\acute{v}_2$. By proposition (\ref{thm:agreement-stability}) agreement then continuous to hold in these pattern, which implies that they are disjoint. Moreover $v_1$ can not be in the past of $\acute{v}_2$ and vice versa. Hence the entire history must be disjoint and therefore any round between $u_1=min\{s_1,\acute{s}_1\}$ must have disjoint last vertices. This however implies that any Lamport graph $G(t)$ has voting weight $\sum_{j=s}^{u_1} w_j^{G(t)}/d_j> 6$ for all $t\geq t_2$. 

However since $\textsc{leader}_{G(t)}(r)$ does not converge by assumption we can repeat the argument an unbounded amount of times, which implies $\sum_{j=s}^{u_i} w_j^{G(t_2)}/d_j> 6$ for arbitrary large round numbers $u_i$ which violates our assumption (\ref{eq:difficulty-bound}) on the difficulty oracle bound.   
\end{proof}

\begin{corollary}[Leader stream convergence]
\label{thm:leader-stream-convergence}
Suppose that the probability for the appearance of new rounds and safe voting pattern is not zero and let $j,k\in \Pi$ be two honest process. Then their leader streams will converge. 
\end{corollary}
\begin{proof}
The previous theorem (\ref{thm:leader-stream-convergence}) implies that each leader set will converge to a single element, which implies that the leader stream of each honest process will converge and it remains to show that the elements in both leader streams are identical. This however follows from our message dissemination assumption (\ref{assumptions}), since Lamport graphs of honest processes eventually converge to contain the same elements. 
\end{proof}

All previous proof are based on the assumption that the voting weight function is essentially unbounded and that the difficulty oracle can be estimated by assumption (\ref{eq:difficulty-bound}) only. However some implementations might be much simpler, in that they don't have unbounded weight function or a strict upper bound on the voting can be computed. As the following corollary shows, this leads to much faster and cleaner convergence of the global leader stream.
\begin{corollary}[Bounded voting weight leader stream]
\label{thm:bounded-weight-decision}Let $d$ be a difficulty oracle, $G$ a Lamport graph and $\textsc{leader}_G(\cdot)$ the leader stream of $G$, such that the overall amount of voting weight in any round $r$ is always in between $3 d_r < w \leq 6 d_r$. Then $\textsc{leader}_G(r)$ will never contain more then one element.
\end{corollary}
\begin{proof}This follows directly from proposition (\ref{thm:each-local-leader-adds-3ds}) as more then one element would imply that there is a round with overall voting weight strictly larger then $6d_r$.
\end{proof}

\subsubsection{Total order convergence}
 We proof that Moser \& Melliar-Smith's properties of a byzantine resistant total order algorithm as defined in (\ref{def:byzantine-total-order-axioms}) are satisfied, provided our set of assumptions (\ref{assumptions}) holds.

\begin{proposition}[Partial Correctness] The asymptotically convergent total orders determined by any two non byzantine processes are consistent; i.e., if any non byzantine process has a Lamport graph that determines $v.total\_position=i$, then no honest process has a Lamport graph that determines $\acute{v}.total\_position=i$, where $\acute{v}\not\equiv v$.
\end{proposition}
\begin{proof}
The theorem follows, since the leader streams of two honest processes eventually converge with probability one, by corollary (\ref{thm:leader-stream-convergence}) and the order is derived deterministically from the past of any element in the leader stream only. However that is equivalent among all Lamport graphs, by the equivalence of the past theorem (\ref{thm:invariant-past}).  
\end{proof}

\begin{proposition}[Consistency] The total order determined by any non byzantine process is consistent with the partial causality order; i.e. $\acute{v}\leq v$ implies $\acute{v}.total\_position\leq v.total\_position$.
\end{proposition}
\begin{proof}
Let $v$ and $\acute{v}$ be two vertices in a Lamport graph with $\acute{v}\leq v$, such that the total order of both vertices is not $\bot$. Let $v$ be in the order cone of some momentrary round leader $v_l$, e.g. $v\in Ord(v_l)$. Then either $\acute{v}\in Ord(v_l)$ or not. In the former case both $v.total\_order$ and $\acute{v}.total\_order$ are computed by a topological sorting algorithm hence $\acute{v}\leq v$ implies $\acute{v}.total\_order\leq v.total\_order$ by the very properties of topologic sorting. In the latter case $\acute{v}\leq v$ implies $\acute{v}\in G_{v_l}$, but since $\acute{v}\not\in Ord(v_l)$, we know $\acute{v}.total\_order < w.total\_order$ for all $w\in Ord(v_l)$, since the total order of all elements from $Ord(v_l)$ starts with a value greater then all previous totally ordered elements.
\end{proof}
\begin{proposition}[Probabilistic Termination I] The probability that a honest process $j$ computes $v.total\_position=i$ for some Lamport graph $G$, position $i$ and vertex $v$ increases asymptotically to unity as the number of steps taken by $j$ tends to infinity.
\end{proposition}
\begin{proof}
By theorem (\ref{thm:leader-stream-convergence}) every round $r$ will converge to a single round leader $l_r$ and at least some of these rounds will converge to a leader $\neq \oslash$ due to our initial vote assumption (\ref{assumptions}). This implies that the order loop (\ref{alg:order-loop}) will execute some topological sorting, like (\ref{alg:kahn-sorting}), that assigns a total order position to all elements in the past of a leader vertex $v$ with $v.m=l_r$ in any Lamport graph. As the leader converges, so does the order in its past and as this goes on forever there will eventually be a round $u$ and a leader $l_u$, such that $|G_{v}|>i$ for $v$ with $v.m=l_u$ and any $i\in\mathbb{N}$. Hence Lamport graph $G$ will have a vertex with $v.total\_order=i$ and this vertex converges to a fixed value.
\end{proof}
\begin{proposition}[Probabilistic Termination II] For each message $m$ broadcast by a non byzantine process $j$, the probability that a non byzantine process $k$ places some vertex $v$ with $v.m=m$ in the total order, increases asymptotically to unity as the number of steps taken by $k$ tends to infinity.
\end{proposition}
\begin{proof}
Let $m$ be a message and $v$ a vertex with $v.m=m$ and round number $r$. Since new rounds appear forever, there will eventually be a round $r$ and round $r$ leader vertices $v_l$ in the future of $v$, (i.e $v\in G_{v_l}$) that converge to a single element. As the past of these leaders gets ordered by algorithm (\ref{alg:kahn-sorting}) and this order converges, the order of $v$ converges too. 
\end{proof}
\section{Conclusion \& Outlook}
We have developed a family of total order algorithms that may survive in unstructured Peer-2-Peers networks and in the presence of momentary large amount of partitioning and faulty behavior. The system chooses different strategies with respect to Brewers CAP-theorem. If no forking occurs, it choose a strict availability \& consistency strategy which allows for short finality. However if partition occurs, incarnated in the form of more then one safe voting pattern in a round, availability remains but consistency becomes probabilistically convergent only.

In contrast to most other approaches, our system is able to incorporate a Proof-Of-Work based voting strategy, which circumvent serious Proof-Of-Stake problems, like bootstrapping and runaways. Proof-Of-Work is difficult to use in byzantine agreement, because such a weight function is usually unbounded.

In any case, future research has to be made in finding optimal system parameter, like the difficulty oracle, incentivation \& punishment, weight systems and the quorum selector. Of course its possible to just be creative and make these function up, but the author believes that a systematical search to find optima is much more reasonable.  

However, if you would like to support the continuous production of content like this, please donate via one of the channels mentioned on the title page, or contact the author for additional solutions.
%===========================================================================================
%						***		References		***
%===========================================================================================
%\newpage

\end{document}